\documentclass[floatfix, aps, pra, amsmath, twocolumn, superscriptaddress, footinbib, longbibliography]{revtex4-1}

\usepackage{amsmath,amssymb}
\usepackage{amscd, latexsym}
\usepackage{mathrsfs}
\usepackage{graphicx}
\usepackage{epstopdf}
\usepackage{cancel}
\usepackage{amsfonts}
\usepackage{exscale}
\usepackage{dcolumn}
\usepackage{bm}
\usepackage{color}
\usepackage{natbib}
\usepackage{soul}

\usepackage{lmodern}
\usepackage[T1]{fontenc}
\usepackage[latin9]{inputenc}
\usepackage{verbatim}
\usepackage{float}
\usepackage{hyperref}
\usepackage{breakurl}

\newcommand{\ha}{\hat{a}}

\begin{document}

\title{Counting statistics of microwave photons in circuit QED}

\author{Konstantin N. Nesterov}
\affiliation{Department of Physics, University of Wisconsin-Madison, Madison, WI 53706}
\author{Ivan V. Pechenezhskiy}
\affiliation{Department of Physics, Joint Quantum Institute,
and 
Center for Nanophysics and
\\ Advanced Materials,
University of Maryland, College Park, MD 20742}
\author{Maxim G. Vavilov}
\affiliation{Department of Physics, University of Wisconsin-Madison, Madison, WI 53706}

\date{\today}

\begin{abstract}
In superconducting circuit architectures for quantum computing, microwave
resonators are often used both to isolate qubits from the electromagnetic environment and to facilitate qubit state readout. We analyze the full counting statistics of photons emitted from such driven readout resonators  both in and beyond the dispersive approximation. We calculate the overlap between emitted-photon distributions for the two qubit states and explore strategies for its minimization with the purpose of increasing fidelity of intensity-sensitive readout techniques. In the dispersive approximation and at negligible qubit relaxation, both distributions are Poissonian, and the overlap between them can be easily made arbitrarily small. Nondispersive terms of the Hamiltonian generate squeezing and the Purcell decay with the latter effect giving the dominant contribution to the overlap between two distributions.


\end{abstract}

\maketitle

\section{Introduction}
Architectures based on superconducting qubits have been exceedingly successful in the contest of building a quantum processor~\cite{Devoret2013, Arute2019}. However, fast and high-fidelity single-shot readout of qubit states, which is indispensable for  quantum error corrections~\cite{Bravyi1998, Fowler2012}, remains somewhat challenging in these architectures. 
The best existing readout techniques in superconducting systems are based on the measurement of the transmitted or reflected microwave field from a resonator weakly coupled to the qubit~\cite{Blais2004, Wallraff2005} with fidelities only around 99\%~\cite{Jeffrey2014, Krantz2016, Walter2017}. 
This number is barely at the required minimum surface-code threshold of 99\%--99.5\%~\cite{Martinis2015}. Further improvement of the fidelity requires a more comprehensive understanding of the readout process. 

\begin{figure}[t]
	\includegraphics[width=0.92\columnwidth]{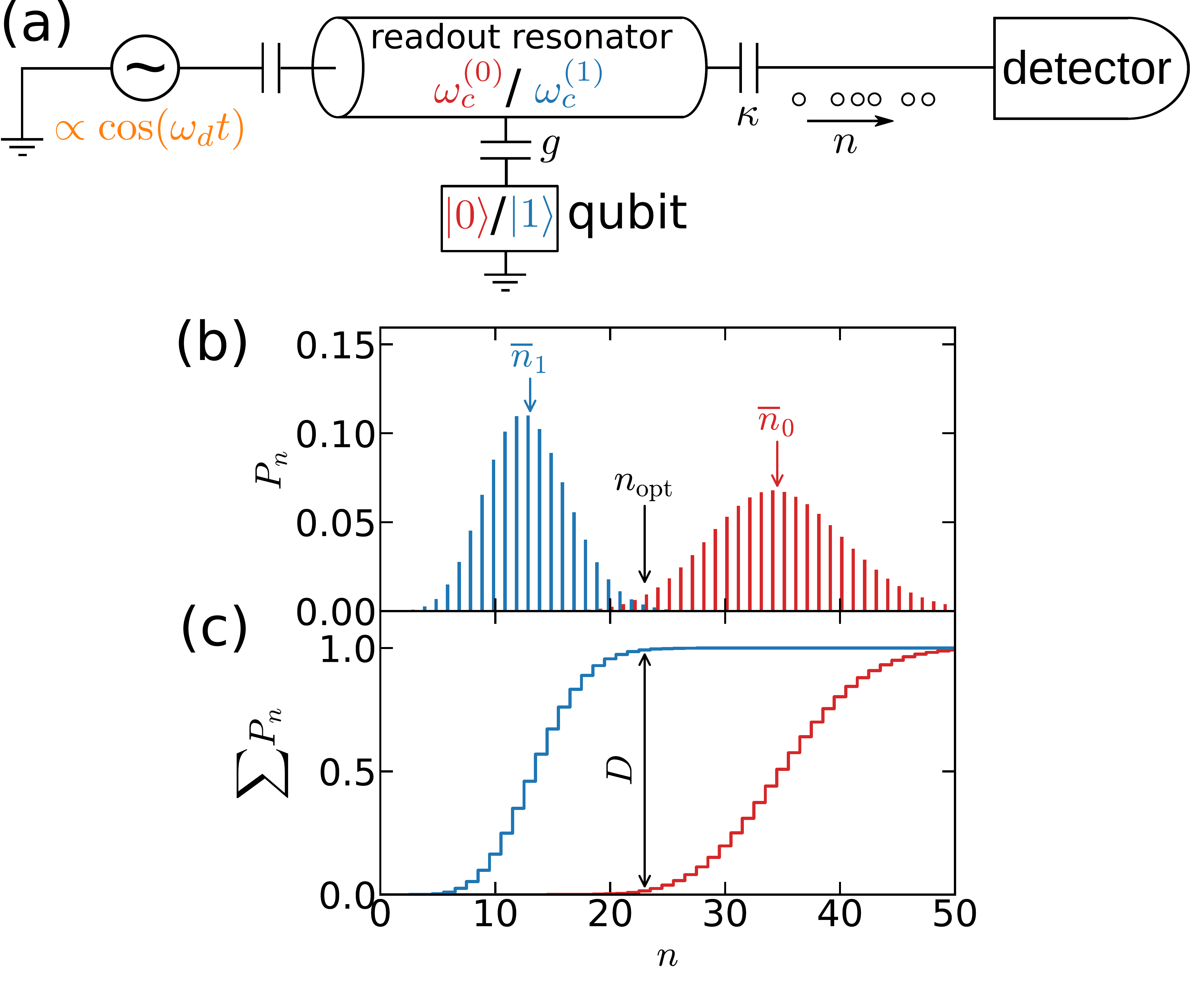}
	\caption{(a) Schematic setup of the circuit QED architecture. The resonator frequency [$\omega_c^{(0)}$ or $\omega_c^{(1)}$] depends on the qubit state. (b), (c) Distributions $P_n$ (b) and integrated distributions $\sum_{n' < n} P_{n'}$ (c) of photons emitted by the cavity for the two qubit states. The Kolmogorov-Smirnov distance $D$ between two distributions is the maximum vertical distance between two integrated distributions. It coincides with the measurement fidelity of a perfect photon counter with thresholding.}\label{fig-definitions}
\end{figure}

In this paper, we present a theoretical analysis of the full counting statistics of photons~\cite{Mandel1963, Kelley1964, Carmichael2008_book, Brange2019} emitted by the readout cavity in a circuit quantum electrodynamics (QED) setup [Fig.~\ref{fig-definitions}(a)]. The knowledge of these statistics is important  because the particle nature of light results in photon shot noise, which affects qubit dynamics and fundamentally limits the measurement rate~\cite{Clerk2010}. In this paper, we focus on analyzing distinct features of the photon distributions for the two qubit states [Fig.~\ref{fig-definitions}(b)]. We parametrize and calculate the overlap between these distributions, explore strategies of its minimization, and address the feasibility of identifying qubit states by counting emitted photons.

Readout techniques that involve photon counting are standard in atomic systems~\cite{Myerson2008, Bochmann2010, Gehr2010, Kwon2017}, where the resonance fluorescence spectroscopy demonstrates single-shot readout fidelities exceeding 99.9\% for qubits based on neutral atoms~\cite{Gehr2010} and 99.99\% for optical qubits stored in trapped ions~\cite{Myerson2008}. In circuit QED, a single-shot qubit measurement using a detector of microwave photons has  been recently demonstrated in Ref.~\cite{Opremcak2018}.
A key advantage of this technique  is that the measurement signal does not require postprocessing outside of a dilution refrigerator. 
When integrated with single-flux-quantum control of qubits~\cite{McDermott2018, Leonard2019, Li2019}, this approach can provide a scalable platform for low-latency feedback, which is necessary for an active error correction. In contrast, the conventional qubit readout approach in circuit QED~\cite{Blais2004, Wallraff2005} is  hardly scalable  as it uses bulky nonreciprocal elements inside dilution refrigerators and room-temperature heterodyne detection and thresholding.

Possible implementations of microwave-photon detectors include those  based on a lossy flux qubit~\cite{Opremcak2018}, on a current-biased Josephson junction (CBJJ)~\cite{Romero2009_prl, Romero2009_physscr, Peropadre2011, Chen2011, Poudel2012, Govia2012, Govia2014, Schoendorf2018_qst}, 
and on a semiconductor double quantum dot~\cite{Wong2017}. The first two types are frequently referred to as the Josephson photon multipliers (JPMs). The efficiency and optimal working conditions of the  JPM have been studied in several theory papers~\cite{Peropadre2011, Poudel2012, Govia2012, Govia2014, Schoendorf2018_qst}. There have also been  proposals on how to use JPMs to perform parity measurements on multiple qubits~\cite{Govia2015, Schoendorf2018_pra}.  
 Experimentally, the flux-qubit type of the JPM has shown qubit readout fidelities of 92\%~\cite{Opremcak2018}, and the CBJJ-based JPM has been used to measure the coincidence counting statistics of microwave photons~\cite{Chen2011}. In addition, a CBJJ-based detector enables  classical-microwave-radiation detection with a single-photon strength~\cite{Oelsner2017}. A more comprehensive review of microwave-photon detection can be found in Ref.~\cite{Gu2017}.


In connection to qubit readout, our theoretical analysis of emitted-photon statistics gives quantitative predictions for the accuracy of identifying a qubit state with a perfect counter of microwave photons. We quantify the overlap between two  distributions by the  Kolmogorov-Smirnov distance [Fig.~\ref{fig-definitions}(c)], which is exactly the measurement fidelity of an ideal counter with thresholding. To improve the distinguishability between the distributions,  we consider mapping of the two qubit states into  cavity states with substantially distinct photon numbers that we refer to as  the bright and dark states~\cite{Govia2014}. The preparation of such  bright (large photon number) and dark (small photon number) cavity states is an essential ingredient of the readout technique of Ref.~\cite{Opremcak2018}. 

Throughout the paper, we use the term ``counting'' for brevity when referring to a specific time interval for which we calculate the photon statistics. We discuss two protocols of photon counting. In the continuous protocol, we calculate the statistics of photons emitted during the application of a microwave drive. In the sequential protocol, we separate in time the preparation of cavity  states and photon counting, as proposed in Ref.~\cite{Govia2014}. In this case, we analyze the statistics of photons emitted by the cavity after the drive is turned off.



We first consider the limit of a large frequency detuning and weak coupling between the qubit and resonator, when the system is well described by the dispersive Hamiltonian~\cite{Blais2004}. To address theoretical bounds on the overlap between two photon distributions and to simplify analytic treatment, we ignore any qubit relaxation. Under these assumptions, the statistics of photon emission are Poissonian and the distributions are most distinguishable when the intensities of the emitted radiation for the two qubit states have highest contrast. The overlap decreases monotonically and can be made arbitrarily small by increasing the microwave-drive power or, for the continuous protocol, by increasing the counting time. For the same counting time and maximum cavity occupation, the sequential protocol gives  a smaller value of the threshold that is needed for determining which distribution a specific number of counted photons belongs to. A  continuous protocol in the dispersive approximation has been previously studied in Ref.~\cite{Sokolov2016} for steady states of the cavity; here we consider the time evolution of  cavity states and focus on shorter time scales.

We next analyze the statistics of emitted photons when the dispersive approximation is no longer valid, performing simulations for realistic parameters of coupled transmon qubit~\cite{Koch2007} and cavity. In this case, the nonlinear resonator shows some degree of squeezing~\cite{Khezri2016}, which, in turn, reduces the width of the emitted-photon distribution function. At the same time, the Purcell decay~\cite{Blais2004,  Sete2014} 
results in widening and skewing of this distribution for the qubit excited state and reduces accuracy in identifying qubit states. We demonstrate that these effects of qubit relaxation are smaller in the sequential protocol.
In addition, we find that increasing the drive power is no longer efficient in reducing the overlap between distributions  and that this overlap no longer decreases monotonically as a function of counting time, but has a minimum.  
For calculations both in and beyond the dispersive limit, we show that the cavity decay rate that minimizes the overlap is parametrically smaller than the decay rate that optimizes the signal-to-noise ratio in a traditional heterodyne technique~\cite{Gambetta2008}.

The outline of the paper is as follows. In Sec.~\ref{sec-motivation}, we define a metric characterizing the distinguishability between two photon distributions, introduce our notation, and discuss cavity pointer states.
In Sec.~\ref{sec-dispersive}, we study the statistics of emitted photons using an idealized dispersive model.
In Sec.~\ref{sec-realistic}, we go beyond the dispersive approximation and perform our calculations for  transmon qubits using the full-counting-statistics formalism. 
  We conclude in Sec.~\ref{sec-conclusions}.

\section{Concepts and definitions}\label{sec-motivation}

Here we introduce the main concepts  for our full-counting-statistics analysis in the context of qubit measurements. In Sec.~\ref{sec-definitions}, we  discuss the distributions of photons emitted by the cavity and define the distance between such distributions for the two qubit states. 
In Sec.~\ref{sec-htrdn-vs-jpm}, we   compare cavity pointer states for conventional heterodyne readout technique and for the technique based on  microwave-photon detection. 

\subsection{Photon statistics}\label{sec-definitions}

A minimal model of the circuit QED architecture~\cite{Blais2004, Wallraff2004} includes a microwave resonator that is coupled to a qubit and to an outgoing transmission line as shown schematically in Fig.~\ref{fig-definitions}(a). The dispersive readout  relies on the dependence of the cavity resonance frequency [$\omega_c^{(0)}$ or $\omega_c^{(1)}$] on the qubit eigenstate ($|0\rangle$ or $|1\rangle$). Thus, when the resonator is driven by a classical microwave drive,  
$|0\rangle$ and $|1\rangle$ qubit states are mapped into different cavity  states. It is common to refer to these photonic states, which contain the measurement record, as to pointer states with the terminology borrowed from  quantum measurement theory~\cite{Zurek1981}. The goal of this paper is to calculate and analyze the statistics of the number of photons emitted by  pointer states into the transmission line. Motivated by qubit measurements,  we focus on distinct features of these statistics for the two qubit states and, in particular, on how distinguishable the two distributions are.

We will use subscripts and superscripts 0 and 1 for various quantities corresponding to the initial qubit states $|0\rangle$ and $|1\rangle$ at time $t=0$. 
Let  $P_n^{0}$ and $P_n^{1}$ be the probabilities of having exactly $n$ photons emitted by the corresponding cavity pointer states. An example of two such distributions  is shown in Fig.~\ref{fig-definitions}(b). These distributions always have a finite overlap since, at least, there is  a probability of zero photon emission from both states. To quantify such overlap, we first introduce $n$-dependent difference between two integrated probability distribution functions:
\begin{equation}\label{distance-threshold}
 D_n = \left|\sum_{n'<n}\left(P_{n'}^{0} - P_{n'}^{1}\right)\right|\,.
\end{equation}
The integrated probability distribution functions are shown in Fig.~\ref{fig-definitions}(c) for the same distributions as in Fig.~\ref{fig-definitions}(b), so $D_n$ is the vertical distance between them at a specific $n$. The maximum value of this difference gives 
the  Kolmogorov-Smirnov distance between $P_n^{0}$ and $P_n^{1}$:
\begin{equation}\label{KS-definition}
 D = \max_{n} D_n = \max_{n}\left|\sum_{n'<n}\left(P_{n'}^{0} - P_{n'}^{1}\right)\right|\,.
\end{equation}
This maximum distance is shown by a vertical arrow in Fig.~\ref{fig-definitions}(c), and the corresponding
optimal value of $n = n_{\rm opt}$, which maximizes $D_n$, is indicated in Fig.~\ref{fig-definitions}(b). In this paper, we further refer to $1-D$  as the overlap or optimal overlap between $P_n^{0}$ and $P_n^{1}$, to $1-D_n$ -- as the $n$-dependent overlap, and to $n_{\rm opt}$ -- as the optimal separation threshold. 

We now briefly explain how the distances (\ref{distance-threshold}) and (\ref{KS-definition}) are related to the fidelity of qubit readout with a photon detector. 
Here we define such fidelity according to~\cite{Gambetta2007, Walter2017}
\begin{equation}\label{fidelity-definition}
 F = 1 - P_{\rm det}(0|1) - P_{\rm det}(1|0)\,,
\end{equation}
where $P_{\rm det}(0|1)$ is the conditional probability to mistakenly detect $|0\rangle$ for the qubit prepared in $|1\rangle$ and vice versa. Let us assume a detector with the binary (``click'' or ``no click'') output that is insensitive to any information about emitted photons except their total number $n$. In the single-shot measurement, the ultimate role of such a detector is to determine whether a specific value of $n$ belongs to $P_n^0$ or $P_n^1$.
Let  $G_n$ be the probability of detector ``clicking''  conditioned on having exactly $n$ photons emitted. Ideally, $G_0 = 0$ (the detector never ``clicks'' without photons) and $G_n \to 1$ at large $n$ (the detector always ``clicks'' if there are many photons). Then, associating a ``click'' with the distribution with the larger mean value $\overline{n}$, we find for the case shown in Fig.~\ref{fig-definitions}(b)
\begin{eqnarray}
 P_{\rm det}({0 |1}) &=& \sum_n G_n P_n^{ 1}\,,\\
 P_{\rm det}({1 |0}) &=& \sum_n (1-G_n) P_n^{0}\,,
\end{eqnarray}
so
\begin{equation}\label{fidelity-general}
 F = \sum_n (1- G_n )\left(P_n^{1} - P_n^{0}\right)\,.
\end{equation}

For a threshold detector, which ``clicks'' if and only if $n$ is at least a certain threshold value $n_{\rm th}$, $G_n$ is given by the Heaviside step function:
\begin{equation}\label{detector-probability}
G_n = \theta(n - n_{\rm th}) 
=
\begin{cases}
 0 \quad {\rm if}\quad n < n_{\rm th}\,, \\
 1 \quad {\rm if}\quad n\ge n_{\rm th}\,.
\end{cases}
\end{equation}
This provides the threshold-dependent fidelity $F_{n_{\rm th}}$, which is exactly the $n$-dependent distance $D_{n}$ (\ref{distance-threshold}) at $n = n_{\rm th}$. 
At the optimal value of the threshold $n_{\rm th} = n_{\rm opt}$, when $F_{n_{\rm th}}$ is maximal, it coincides with the Kolmogorov-Smirnov distance (\ref{KS-definition}). A perfect threshold detector with the ability to tune $n_{\rm th}$ can be made using a perfect photon counter with thresholding. In this case, a binary output is assigned based on whether $n$, the number of counted photons, exceeds $n_{\rm th}$ or not. 
Therefore, the Kolmogorov-Smirnov distance (\ref{KS-definition}) is the fidelity of qubit readout with  a perfect counter with thresholding.  We note that the use of additional information about photons such as arrival times to the counter can further improve the fidelity~\cite{Myerson2008}.


\subsection{Cavity pointer states}\label{sec-htrdn-vs-jpm}

Here we discuss cavity pointer states~\cite{Zurek1981} that are typical for heterodyne~\cite{Blais2004, Wallraff2005} and for the photon-detector-based~\cite{Opremcak2018} approaches to qubit readout.
The most important parameters that predict the efficiency of the readout at low cavity occupations are the cavity decay rate $\kappa$ and the dispersive shift $\chi$, which we define as
\begin{equation}\label{chi-definition}
2\chi = \omega_c^{(0)} - \omega_c^{(1)}\,.
\end{equation}
We will argue that the optimal relation between $\kappa$ and $\chi$ is different for the two approaches.

We use $N$ to label the number of photons in the cavity and  the following notation to characterize the cavity state:
\begin{equation}
\alpha = \langle \hat{a} \rangle\quad\text{and}\quad \overline{N} = \langle \hat{a}^\dagger \hat{a}\rangle\,.
\end{equation}
Here $\ha$ is the annihilation operator for the relevant cavity mode, and we always work in the frame rotating with the frequency of the microwave drive $\omega_d$, so time-dependent $\alpha(t)$ does not have a rapidly oscillating phase.
For a coherent state of the cavity, the average photon number is related to $\alpha$ as $\overline{N} = |\alpha|^2$, which does not hold in a general situation.

\begin{figure}[t]
 \includegraphics[width=0.95\columnwidth]{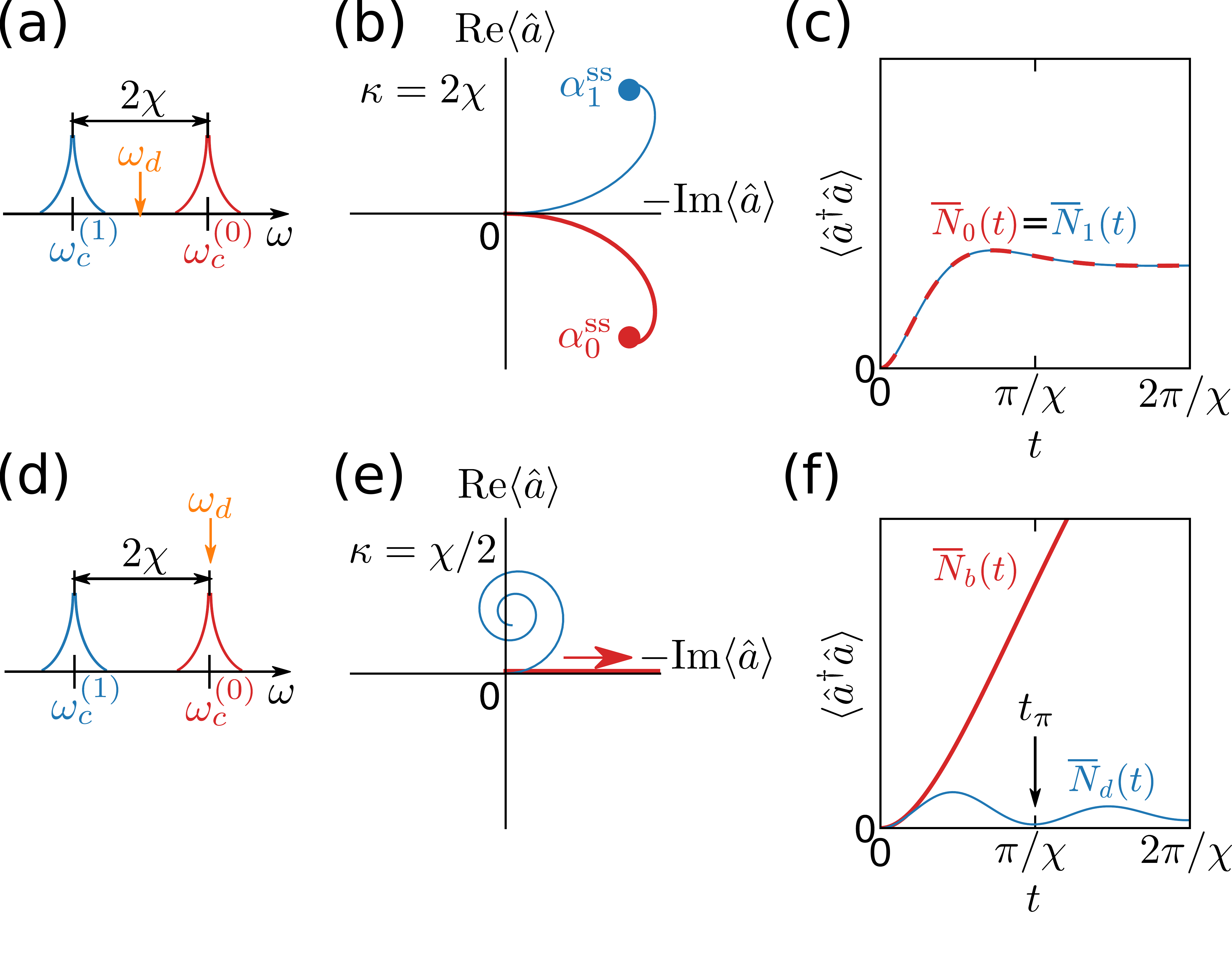}
 \caption{Cavity pointer states in the dispersive regime of circuit QED with negligible qubit relaxation.  (a)--(c) In the ``traditional'' approach, the drive frequency $\omega_d$ is in the middle between $\omega_c^{(0)}$ and $\omega_c^{(1)}$, two qubit-state-dependent trajectories in the quadrature space (b) are symmetric, and the cavity occupation (c) is independent of the qubit state. For the optimal decay rate $\kappa = 2\chi$, the cavity quickly reaches its steady state shown by solid circles in panel (b). (d)--(f) In the photon-detector-based readout, the microwave drive is applied in resonance with one of the dressed cavity frequencies, resulting in the bright and dark states of the cavity with distinct occupations $\overline{N}_b(t)$ and $\overline{N}_d(t)$ for the two qubit states. To reduce  $\overline{N}_d$, it is advantageous to choose a smaller $\kappa/\chi$.}\label{fig-htrdn-vs-jpm}
\end{figure}

In the conventional homodyne and heterodyne techniques, when the reflected or transmitted wave is measured, the microwave drive is often applied   at the frequency $\omega_d$ that is in the middle between  $\omega_c^{(0)}$ and $\omega_c^{(1)}$ [Fig.~\ref{fig-htrdn-vs-jpm}(a)]. In the simplified dispersive model and at negligible qubit relaxation, this leads to symmetric trajectories for the two pointer states in the cavity phase space [Fig.~\ref{fig-htrdn-vs-jpm}(b)] and to the cavity photon occupation that is independent of the qubit state [Fig.~\ref{fig-htrdn-vs-jpm}(c)]; see Sec.~\ref{sec-dispersive} for details of analytic calculations. 
In this case, the relevant characteristic of the cavity state is $\alpha(t)$. The measurement signal (before amplification) is related to the time integral~\cite{Clerk2010}
\begin{equation}\label{heterodyne-signal}
 S(t_0, t) = \sqrt{\kappa} \int \limits_{t_0}^t dt' \alpha(t')\,.
\end{equation}
This is a complex number, which can be reduced to a real number by choosing a specific quadrature.  Typically, $1 / \kappa \ll (t-t_0) < T_1$, where $T_1$ is the qubit lifetime, so this integral is mostly determined by $\alpha^{\rm ss}$, the quasi-steady-state value of $\alpha(t)$. These values  are shown by circles in Fig.~\ref{fig-htrdn-vs-jpm}(b). The measurement rate for an ideal detector efficiency is given by $\Gamma_m = \kappa |\alpha_0^{\rm ss} - \alpha_1^{\rm ss}|^2$. Optimization of this rate  at a fixed steady-state photon occupation $\overline{N}$ results in the optimal cavity decay rate $\kappa = 2\chi$~\cite{Gambetta2008}. This decay rate is relatively large, so the steady states are reached fast, as is evident from Figs.~\ref{fig-htrdn-vs-jpm}(b) and \ref{fig-htrdn-vs-jpm}(c), which are shown for $\kappa =2\chi$. 

While generalized complex-valued signals (\ref{heterodyne-signal}) can be well separated for the two qubit states for this choice of $\omega_d$, the distributions $P_n^0$ and $P_n^1$ overlap strongly and are exactly the same for the approximation used in Fig.~\ref{fig-htrdn-vs-jpm}. Evidently, information about the qubit state is contained in the phase of $\alpha$ rather than its amplitude. 
In measurements using photon detectors, the pointer states are instead created by choosing $\omega_d$ to be in resonance with one of the dressed cavity frequencies, e.g., $\omega_d = \omega_c^{(0)}$ [Fig.~\ref{fig-htrdn-vs-jpm}(d)]. 
In this case, one creates  bright and dark cavity pointer states with distinct photon occupations for the two qubit states [Fig.~\ref{fig-htrdn-vs-jpm}(f)]~\cite{Govia2014, Opremcak2018}. From now on, we will use subscripts and superscripts $b$ and $d$ when describing the system in terms of these bright and dark cavity states. In later sections, we calculate  the corresponding photon distributions $P_n^b$ and $P_n^d$.

In the dispersive limit, the bright-state trajectory $\alpha_b(t)$ in the quadrature space is a straight line, and the dark-state trajectory $\alpha_d(t)$ is a spiral [Fig.~\ref{fig-htrdn-vs-jpm}(e)]. Even though one may choose to prepare similar pointer states for a traditional heterodyne readout, the choice of pointer states with $\overline{N}_0(t) \ne \overline{N}_1(t)$ is essential for a photon detector.
In comparison to the heterodyne readout, the phase of $\alpha$ no longer carries any useful information, and 
the relevant characteristics of a cavity state are now the intensity $|\alpha(t)|$ and the cavity occupation $\overline{N}(t)$. 
The mean number of photons emitted into the transmission line in the time interval between $t_0$ and $t$ is given by~\cite{Clerk2010}
\begin{equation}\label{nbar-definition}
  \overline{n}(t_0, t) = \kappa\int\limits_{t_0}^{t} \overline{N}(t')dt'\,.
\end{equation}
This is the analog of the measurement signal (\ref{heterodyne-signal}). We note that $\overline{n}$ is precisely the area under the $\overline{N}$ vs $t$ curve multiplied by $\kappa$. 

In this paper, we will argue that, for the photon-detector-based measurement, it is justifiable to choose a smaller value of $\kappa/\chi$ in comparison to the optimal value of $2$ found for the conventional readout. Thus we used $\kappa/\chi = 0.5$ in Figs.~\ref{fig-htrdn-vs-jpm}(e) and \ref{fig-htrdn-vs-jpm}(f).  In this case, $\overline{N}_d(t)$ displays decaying oscillations with  minima at time points that are close to integer multiples of 
\begin{equation}\label{t_pi}
t_{\pi} = \frac{\pi}{\chi}\,.
\end{equation}
When $\kappa = 0$, those minima are exactly at multiples of $t_\pi$, the corresponding minimal values of $\overline{N}_b(t)$ are exactly zeros, and the dark-state phase-space trajectory is a circle~\cite{Govia2014}. 

Given the definition of $\chi$ (\ref{chi-definition}), the meaning of $t_\pi$ is the minimum time that is necessary to distinguish between two dressed resonances $\omega_c^{(0)}$ and $\omega_c^{(1)}$. Since both the conventional heterodyne technique and the photon-detector approach ultimately rely on distinguishing between $\omega_c^{(0)}$ and $\omega_c^{(1)}$, the measurement time cannot be fundamentally shorter than $t_\pi$ because of the            time-frequency uncertainty relation. For the photon-detector approach, $t_\pi$ is the shortest time required to  prepare a high contrast between bright and dark pointer states with a minimal $\overline{N}_d$. 
Thus, 
in the sequential protocol of photon counting, which is discussed below in Secs.~\ref{sec-dispersive-onoff} and \ref{sec-realistic-onoff}, the microwave drive is applied only at $0 < t < t_\pi$, while photon counting starts at $t=t_\pi$ to ensure small $\overline{n}_d$.

\section{Dispersive approximation}\label{sec-dispersive}

In this section,  we assume a perfectly linear cavity so that its dressed frequencies $\omega_c^{(0)}$ and $\omega_c^{(1)}$ are independent of the cavity occupation or the drive power. We also ignore the Purcell effect and any other kind of qubit relaxation. These assumptions allow us to focus on contributions to the distributions overlap $1-D$ that are intrinsic for a specific counting and pointer-states-preparation protocol and, therefore, to address theoretical limits on $1-D$. Here we have only two parameters describing the qubit-cavity system: $\chi$ and $\kappa$. Under these assumptions, the cavity state is always a coherent state, and any photon distribution is always Poissonian. 

In Sec.~\ref{sec-dispersive-model}, we introduce the dispersive model~\cite{Blais2004} and explain how the distributions $P^b_n$ and $P^d_n$ can be calculated analytically. In Sec.~\ref{sec-dispersive-alwon}, we discuss the continuous protocol, where photon counting starts from the moment the microwave drive is turned on, and the drive remains on while  photons are counted. Finally, in Sec.~\ref{sec-dispersive-onoff}, we discuss the sequential protocol, where the preparation of pointer states and photon counting are separated in time.

\subsection{Model}\label{sec-dispersive-model}

In the dispersive approximation, the cavity and qubit dynamics are decoupled, and the Hamiltonian of the driven qubit-cavity system is block-diagonal in qubit indices. For a given qubit state ($|0\rangle$ or $|1\rangle$), its block in the rotating frame and in the rotating-wave approximation can be written as~\cite{Blais2004}
\begin{equation}\label{Hamiltonian-disp}
 \hat{H}^{(0/1)} = \hbar\Delta_{\rm cd}^{(0/1)} \left(\hat{a}^\dagger \hat{a} + 1/2\right) + \hbar\varepsilon(t)\left(\hat{a}  + \hat{a}^\dagger \right)\,.
\end{equation}
Here, the difference between the dressed cavity and drive frequencies is given by
\begin{equation}\label{detuning-cavity-drive}
 \Delta_{\rm cd}^{(0/1)} = \omega_c^{(0/1)} - \omega_d\,,
\end{equation}
and $\varepsilon$ is the amplitude of the classical microwave drive. We have omitted qubit energy contribution, which is a constant term in each block (\ref{Hamiltonian-disp}).
In the presence of cavity decay with the rate $\kappa$, this system is described by the following  semiclassical equation of motion 
for the cavity coherent state $|\alpha\rangle$~\cite{Gambetta2008}:
\begin{equation}\label{alpha-semiclassical-eom}
 \dot{\alpha} = -i\Delta_{\rm cd}^{(0/1)}\alpha - \frac{\kappa}{2}\alpha - i\varepsilon(t)\,.
\end{equation}

For $\omega_c^{(0)} > \omega_c^{(1)}$ and the choice $\omega_d = \omega_c^{(0)}$, so qubit in $|0\rangle$ corresponds to the bright cavity state, we find for the bright and dark states for a constant drive [$\varepsilon(t) = \varepsilon = {\rm const.}$] and the cavity originally prepared in its vacuum state [$\alpha(0)=0$]:
\begin{equation}\label{alpha-t-disp}
\left\{
\begin{array}{l}\displaystyle 
 \alpha_b(t) = \frac{2i\varepsilon}{\kappa}\left[\exp\left(-\frac{\kappa t}{2}\right) - 1\right] \,,\\
 \displaystyle
 \alpha_d(t) = \frac{2i\varepsilon}{\kappa - 4i\chi}\left[\exp\left(2i\chi t-\frac{\kappa t}{2}\right) - 1\right]\,.
\end{array}
\right.
\end{equation}
These equations explain the forms of the bright and dark-state trajectories in the phase space, which are shown in Fig.~\ref{fig-htrdn-vs-jpm}(e).
Since $\overline{N} = |\alpha|^2$ for coherent states, Eq.~(\ref{alpha-t-disp}) results in the following time-dependent cavity occupations:
\begin{equation}\label{ncav-t-disp}
\left\{
\begin{array}{l}
 \displaystyle
 \overline{N}_b(t) = \frac{4\varepsilon^2}{\kappa^2} \left[\exp\left(-\frac{\kappa t}{2}\right) - 1\right]^2\,, \\
  \displaystyle
 \overline{N}_d(t) = \frac{4\varepsilon^2}{\kappa^2 + 16\chi^2} \left|\exp\left(2i\chi t-\frac{\kappa t}{2}\right) - 1\right|^2\,.
\end{array}
\right.
\end{equation}
In particular, at $t =t_\pi= \pi/\chi$, we find 
\begin{subequations}
\begin{equation}\label{n_d-n_b}
\overline{N}_d(t_\pi) = \frac{\kappa^2}{\kappa^2 + 16\chi^2} \overline{N}_b(t_\pi)
\end{equation}
and
\begin{equation}\label{ncav-bright-t-pichi}
\overline{N}_b(t_\pi) = N_\pi = \frac{4\varepsilon^2}{\kappa^2} \left[1 - \exp\left(-\frac{\pi\kappa}{2\chi}\right)\right]^2\,.
\end{equation}
\end{subequations}

Using Eqs.~(\ref{nbar-definition}) and (\ref{ncav-t-disp}), we  calculate the average number of emitted photons  provided we start counting at $t_0=0$. We thus find
\begin{subequations}
\begin{equation}\label{nbar-emitted-b}
\overline{n}_b(0, t) =  \frac{4\varepsilon^2}{\kappa^2}\left[\kappa t + 1 - \left( e^{-\kappa t/2} - 2\right)^2 \right]\,,
\end{equation}
and
\begin{multline}\label{nbar-emitted-d}
\overline{n}_d(0, t) 
= \frac{4\varepsilon^2}{\kappa^2 + 16\chi^2}\left(\kappa t + 1 - e^{-\kappa t} - \frac{4\kappa^2}{\kappa^2 + 16\chi^2} \right.\\  \left.
\times\left\{-1 + \left[\cos(2\chi t) - \frac{4\chi}{\kappa}\sin(2\chi t)\right]e^{-\kappa t/2}\right\}\right)\,.
\end{multline}
\end{subequations}
Since, for a coherent state, the photon number in the cavity is distributed according to the Poisson law, the emitted photons follow the Poisson distribution as well, which has been shown for cavity steady states in Ref.~\cite{Sokolov2016}. We have verified this statement for time-dependent states by a direct numerical computation of $P_n^b$ and $P_n^d$ using the full-counting-statistics formalism (see Sec.~\ref{sec-realistic} and Appendix~\ref{sec-fcs}) applied for the dispersive Hamiltonian. 
Thus, we find ($s=b, d$)
\begin{equation}\label{Pn-disp}
 P_n^{s} = \frac{\overline{n}^n_{s}}{n!} e^{-\overline{n}_{s}}\,,
 \end{equation}
where $\overline{n}_b$ and $\overline{n}_d$ are given by Eqs.~(\ref{nbar-emitted-b}) and (\ref{nbar-emitted-d}). 


\subsection{Continuous counting}\label{sec-dispersive-alwon}

\begin{figure}[t]
 \includegraphics[width=0.95\columnwidth]{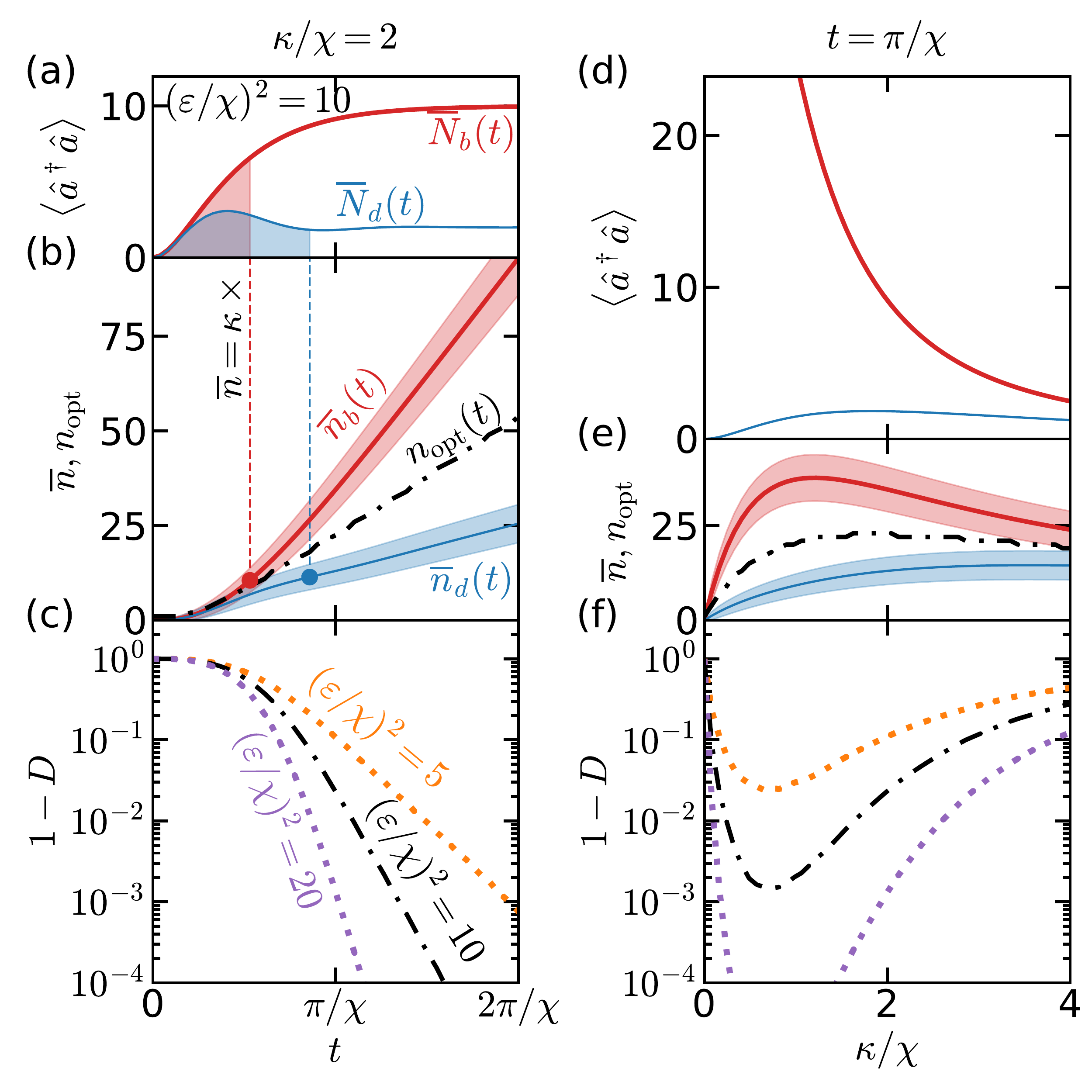}\caption{Continuous photon counting in the dispersive approximation. Photon statistics are shown
 as a function of $t$ at constant $\kappa/\chi =2$ (left column) and as a function of $\kappa/\chi$ at constant $t=\pi/\chi$ (right column). (a), (d) Average cavity occupations for the bright (thick, red) and dark (thin, blue) states at $(\varepsilon/\chi)^2 = 10$. (b), (e) Mean values (solid lines)
 and standard deviations (shaded regions) of emitted-photon distributions for the same parameters as on top panels.
 Vertical lines in (b) and shaded areas in (a) illustrate that $\overline{n}(t)$ is given by the corresponding time integral of $\overline{N}(t)$ up to a factor of $\kappa$; see Eq.~(\ref{nbar-definition}). Dash-dotted lines show the optimal separation threshold $n_{\rm opt}$. (c), (f) Distribution overlaps  for $(\varepsilon/\chi)^2 = 5, 10$, and $20$.}\label{fig-fid-disp-alw-on}
\end{figure}

In this section, we apply the formalism of Sec.~\ref{sec-dispersive-model} to study photon statistics in the dispersive approximation for a constant drive term $\varepsilon(t) = \varepsilon$ and with photon counting starting at $t_0=0$.  We use Eq.~(\ref{Pn-disp}) to find $n = n_{\rm opt}$ that defines the distance $D$ (\ref{KS-definition}) and  calculate the overlap $1-D$ between $P_n^b$ and $P_n^d$. In the analysis of Ref.~\cite{Sokolov2016}, where a similar continuous protocol in the dispersive approximation has been studied, a starting time $t_0 \gg 1/\kappa$ has been assumed to ensure that the cavity is in its steady state.

The cavity occupation $\overline{N}$ and average number of emitted photons $\overline{n}$  for $(\varepsilon/\chi)^2 = 10$ are shown in Figs.~\ref{fig-fid-disp-alw-on}(a) and (b) as a function of time $t$ for the bright (thick red lines) and dark (thin blue lines) states. The standard deviations of $P_n^b$ and $P_n^d$, which are simply $\sqrt{\overline{n}_b}$ and $\sqrt{\overline{n}_d}$ for Poissonian distributions, are illustrated by shaded regions in Fig.~\ref{fig-fid-disp-alw-on}(b). We show the optimal separation $n = n_{\rm opt}$ and the overlap $1 - D$  by dash-dotted lines in panels (b) and (c), respectively.  Here we choose $\kappa = 2\chi$ to relate to conventional dispersive readout techniques, where this value of the decay rate optimizes the signal-to-noise ratio~\cite{Gambetta2008}.

The dependence of both $\overline{N}(t)$ and $\overline{n}(t)$ on $\varepsilon$ is trivial for both bright and dark states:  in the linear approximation for the cavity, they scale with the drive power as $\varepsilon^2$. The behavior of $n_{\rm opt}$ vs $\varepsilon$ is similar except that it scales as $\varepsilon^2$ only approximately because it has discrete values. Since the relative widths of the distributions are simply $1 / \sqrt{\overline{n}}$ and the distributions do not have long tails, the overlap between $P_n^b$ and $P_n^d$ decreases with increasing $\varepsilon$. Therefore, in the approximation under consideration, the dependence of $D$ on $\varepsilon$ is also monotonic, although more complicated than $\overline{n}$ or $\overline{N}$. Thus, in addition to $(\varepsilon/\chi)^2 = 10$, we show $1-D$ in panel (c) for two additional values of the drive power: $(\varepsilon/\chi)^2 = 5$ and $20$. We notice that, for the chosen value of the decay rate, $(\varepsilon/\chi)^2$ is  the bright-state cavity occupation in the steady state, which is evident from Fig.~\ref{fig-fid-disp-alw-on}(a).

 For $(\varepsilon/\chi)^2 = 10$, we observe that the overlap decreases down to $0.01$ at $t \sim t_\pi = \pi/\chi$, which is precisely the minimum time required to resolve two dressed resonator frequencies. For the drive power reduced by a factor of two, the $0.01$ threshold can be reached at $t\sim 1.5 t_\pi$. Overall, $1-D$ decreases very fast with time: for example, waiting $1.5 t_\pi$ instead of $t_\pi$ for $(\varepsilon/\chi)^2 = 10$ reduces $1-D$ by an order of magnitude. 
While increasing the drive power  also reduces the overlap, the time required to achieve the same value of $1-D$ decreases only slightly with increasing $\varepsilon$. For example, $1-D = 0.01$ can be achieved only slightly faster for $(\varepsilon/\chi)^2 = 20$ than for $(\varepsilon/\chi)^2 = 10$. 

In the right column of Fig.~\ref{fig-fid-disp-alw-on}, we demonstrate the cavity-decay-rate dependence of the same metrics and for the same values of $\varepsilon$ as in the left column at a specific time $t=t_\pi$.  At $\kappa =0$, we have $\overline{N}_b(t_\pi) = (\varepsilon t_\pi)^2 = 10\pi^2$ and $\overline{N}_d(t_\pi) =0$ [see Eqs.~(\ref{n_d-n_b}) and (\ref{ncav-bright-t-pichi})]. In this case, the dark-state trajectory in Fig.~\ref{fig-htrdn-vs-jpm}(e) is  a circle, which crosses zero at integer multiples of $t_\pi$~\cite{Govia2014}. With increasing $\kappa$, the bright-state occupation decreases monotonically, while the dark-state occupation displays a local maximum. The difference between $\overline{N}_b(t_\pi)$ and $\overline{N}_d(t_\pi)$ decreases, and both occupation numbers eventually approach zero. Both $\overline{n}_b$ and $\overline{n}_d$  display local maxima with the maximum for the bright state occurring at a smaller $\kappa$.
At $\kappa\ll \chi$, the values of $\overline{n}$ are suppressed since hardly any photons are leaking out of the cavity. At $\kappa\gg \chi$, the photon flux from the cavity is also small since the cavity occupation is small. In other words, a high decay rate prevents the creation of the cavity states with sufficiently large photon occupations, which reduces the number of emitted photons. As a consequence of  the nonmonotonic behavior of $\overline{n}$ for both states, the overlap $1-D$ displays a local minimum at $\kappa/\chi \lesssim 1$; see Fig.~\ref{fig-fid-disp-alw-on}(f).


In general, we observe that an arbitrarily small overlap $1-D$ is straightforward to achieve in this protocol. A small $1-D$ normally implies relatively large values of  $\overline{n}_d$ and $n_{\rm opt}$. As can be seen in Figs.~\ref{fig-fid-disp-alw-on}(b) and \ref{fig-fid-disp-alw-on}(d), we have $\overline{n}_d \gtrsim 10$ and $n_{\rm opt} \gtrsim 20$ for $t\sim t_\pi$ and $1 - D \lesssim 0.01$. In the next section, we discuss a different protocol with a smaller $\overline{n}_d$, which may be useful for an experiment with  a binary photon detector  that is sensitive to a small photon number.


\subsection{Sequential counting}\label{sec-dispersive-onoff}

\begin{figure}[t]
 \includegraphics[width=0.95\columnwidth]{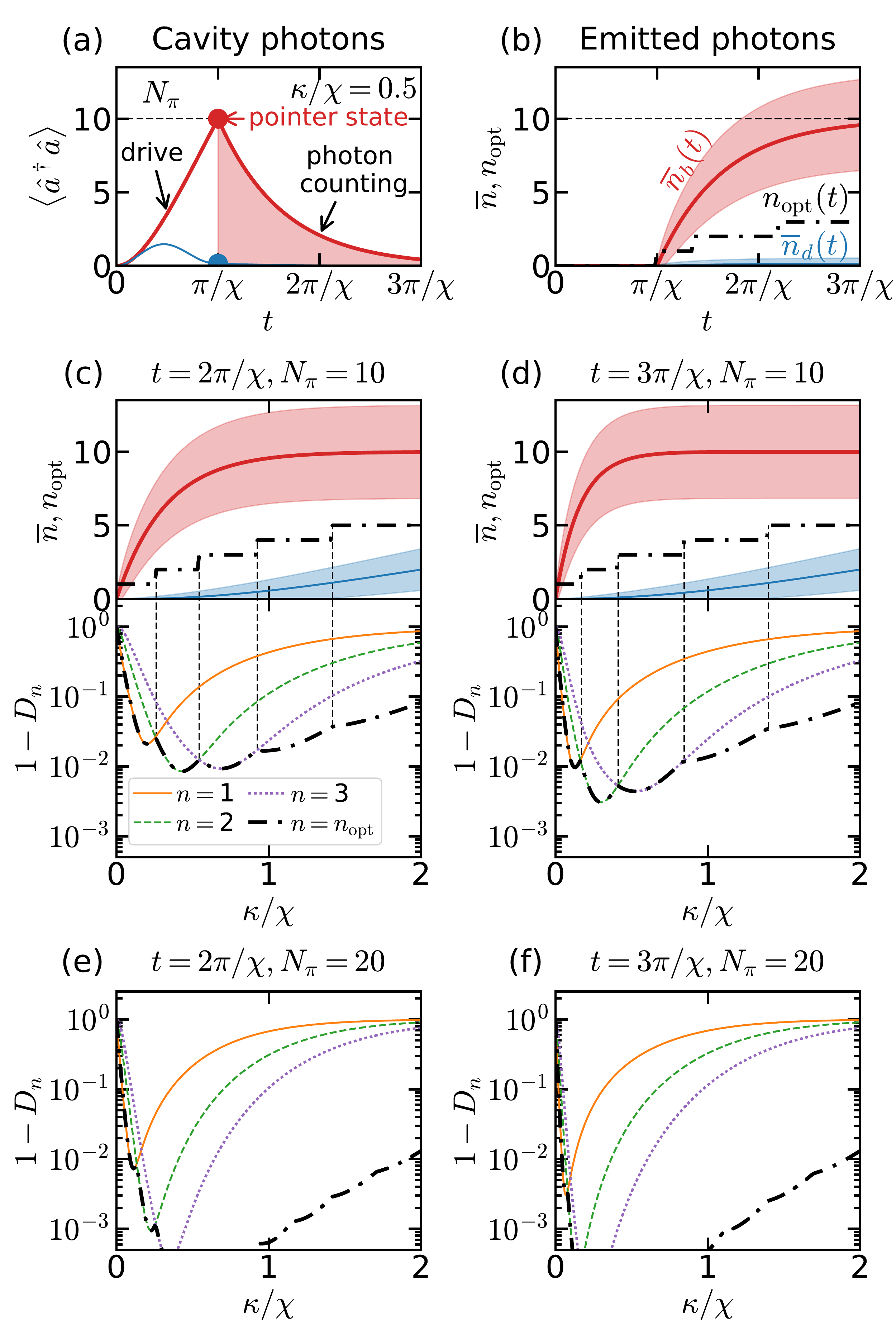}
 \caption{Sequential counting in the dispersive approximation: the microwave drive is applied only between $t = 0$ and $t= t_\pi$ to prepare good cavity pointer states (a), and only photons that are emitted afterwards are counted (b). (c), (d) The statistics of emitted photons as a function of $\kappa$ calculated at $t=2t_\pi$ (c) and $t=3t_\pi$ (d) for the drive amplitude $\varepsilon(\kappa)$ chosen to ensure $N_\pi= \overline{N}_b(t_\pi) = 10$ at each $\kappa$. Top: mean values and standard deviations of $P_n^b$ and $P_n^d$ and the optimal threshold between the distributions. Bottom: $n$-dependent overlap $1-D_n$ for $n= 1$, 2, and 3 photons and the optimal overlap $1-D$, which is labeled as $n=n_{\rm opt}$. (e), (f) Same as in bottom parts of (c) and (d) but for $N_\pi = 20$. Where applicable, line styles and shading follow the convention of Fig.~\ref{fig-fid-disp-alw-on}.
 }\label{fig-fid-disp-onoff}
\end{figure}

Here we discuss a possible way to reduce $\overline{n}_d$ and, consequently, $n_{\rm opt}$ by separating in time the preparation of cavity pointer states and  photon counting. This is similar to the readout protocol of Ref.~\cite{Govia2014}, which was based on a Josephson photon multiplier (JPM), except that we  study the statistics of photons emitted into a transmission line, following the experimental setup of Ref.~\cite{Opremcak2018}. In contrast, a direct capacitive coupling between a JPM and  readout cavity was studied in Ref.~\cite{Govia2014}.
We consider the separation of the two stages at $t = t_\pi$, when the microwave drive is switched off and  photon counting starts. Since this time is around the first local minimum of $\overline{N}_d(t)$ [see Fig.~\ref{fig-htrdn-vs-jpm}(f)], $\overline{n}_d$ is not large and is bounded by $\overline{N}_d(t_\pi)$.

We illustrate this procedure in time domain for the cavity state in Fig.~\ref{fig-fid-disp-onoff}(a)  and, for the distributions of emitted photons, in Fig.~\ref{fig-fid-disp-onoff}(b). The cavity pointer states prepared at $t=t_\pi$ are shown by circles. At $t \le t_\pi$, we find $\overline{N}(t)$ from Eq.~(\ref{ncav-t-disp}), as before, while $\overline{n}(t) = 0$. At $t > t_\pi$, we have $\overline{N}(t) = \overline{N}(t_\pi) \exp[-\kappa (t-t_\pi)]$ and $\overline{n}(t) = \overline{N}(t_\pi) - \overline{N}(t)$. For both bright and dark states, $\overline{n}(t)$ approaches $\overline{N}(t_\pi)$ in the limit $t \rightarrow \infty$, which is shown for the bright state by the horizontal dashed line in Fig.~\ref{fig-fid-disp-onoff}(b). 
As is evident from these figures, only a small number of photons can leak from the dark state after $t=t_\pi$, since $\overline{N}_d(t_\pi)$ is small.

We first estimate the smallest possible  $N_\pi = \overline{N}_b(t_\pi)$ to obtain a desired overlap. According to Eq.~(\ref{n_d-n_b}), for a fixed $N_\pi$, the value of $\overline{N}_d(t_\pi)$ decreases with decreasing $\kappa$ and is exactly zero at $\kappa =0$. If there are no restrictions on the counting time $t_{\rm count}$, the overlap for a fixed $N_\pi$ is minimized when $\overline{N}_d(t_\pi) \rightarrow 0$ and $t_{\rm count} \gg 1/\kappa$ for all the photons from the bright state to leak out of the cavity. 
In this case, $1-D$ is determined by the probability to have zero photons in the bright state, which, for the Poissonian statistics, is simply $\exp(-N_\pi)$.
Therefore, to ensure ${1-D \leq 0.01}$, the bright-state cavity occupation has to be at least $N_\pi = -\ln(0.01)\approx 4.6$, while this value increases up to 6.9  to satisfy $1 - D \leq 0.001$. 

For a counting time that does not substantially exceed $t_\pi$, one can achieve similar conclusions about the smallest $N_\pi$ if a tunable $\kappa$ is implemented in the system. This way, a choice of $\kappa \ll \chi$ at $t < t_\pi$ and $\kappa \gg \chi$ at $t > t_\pi$ would result in an ideal dark pointer state, which is similar to the protocol of Ref.~\cite{Govia2014}, and in a very fast transfer of the bright state towards the photon counter. Here we do not consider such tunable couplers, which can only improve our results, and focus on choosing an optimal time-independent $\kappa$. To demonstrate that a small $1-D$ at a relatively short $t_{\rm count}$ is possible, we show results for larger $N_\pi$.

In this section, when studying various metrics as a function of $\kappa$, we fix the value of $N_\pi$ and adjust the drive amplitude $\varepsilon = \varepsilon(\kappa)$ for each $\kappa$ according to Eq.~(\ref{ncav-bright-t-pichi}). In realistic systems, the strength of nonlinear effects and  breakdown of the dispersive approximation are determined by the cavity occupation rather than by the drive power, so keeping the former constant is the correct way to approach the problem of optimization over $\kappa$.
In Fig.~\ref{fig-fid-disp-onoff}, we show the statistics of emitted photons as a function of $\kappa$ for $N_\pi = 10$ [Figs.~\ref{fig-fid-disp-onoff}(c) and \ref{fig-fid-disp-onoff}(d)] and for $N_\pi = 20$ [Figs.~\ref{fig-fid-disp-onoff}(e) and \ref{fig-fid-disp-onoff}(f)]. To focus on relatively short time scales, we show the results for the end-of-counting time $t$ [the upper limit in Eq.~(\ref{nbar-definition})] to be $2t_\pi$ [Figs.~\ref{fig-fid-disp-onoff}(c) and \ref{fig-fid-disp-onoff}(e)] and $3t_\pi$ [Figs.~\ref{fig-fid-disp-onoff}(d) and \ref{fig-fid-disp-onoff}(f)].

The bright-state $\overline{n}_b$ approaches $N_\pi$ with increasing $\kappa$ [see Figs.~\ref{fig-fid-disp-onoff}(c) and \ref{fig-fid-disp-onoff}(d)] since photons get emitted faster with a higher cavity decay rate.  For the dark state, we find that $\overline{n}_d$ increases and approaches $\overline{n}_b$ since $\overline{N}_d(t_\pi)$ increases and approaches $N_\pi$, see Eq.~(\ref{n_d-n_b}). 
At small $\kappa$, we observe $n_{\rm opt}=1$, which  increases in steps with increasing $\kappa$.
This is demonstrated in bottom parts of Figs.~\ref{fig-fid-disp-onoff}(c) and \ref{fig-fid-disp-onoff}(d) and in Figs.~\ref{fig-fid-disp-onoff}(e) and \ref{fig-fid-disp-onoff}(f), where we show the $n$-dependent overlaps $1 - D_n$ together with the optimal overlap $1-D$ for the $\kappa$-dependent optimal separation threshold $n_{\rm opt}$ [see Eqs.~(\ref{distance-threshold}) and (\ref{KS-definition})]. Vertical dashed lines in Figs.~\ref{fig-fid-disp-onoff}(c) and \ref{fig-fid-disp-onoff}(d) further illustrate the jumps in $n_{\rm opt}$. 

We notice that the values of $\kappa$ minimizing $1-D$ at a fixed $N_\pi$ are  smaller than $2\chi$, which optimizes the signal-to-noise ratio in standard measurement techniques~\cite{Gambetta2008}. 
For $N_\pi = 10$, we see that $1-D$ can drop below 0.01 at $t=2t_\pi$ for $\kappa/\chi \sim 0.5-0.75$ with $n_{\rm opt} = 2$ or 3 photons. For  $n=1$, a longer counting time and a very small $\kappa$ are required to ensure that $1-D_n < 0.01$. Increasing $N_\pi$ generally helps reduce the overlap. Thus, $N_\pi =20$ leads to surpassing of the 0.001 threshold already at $t=2t_\pi$ for $n_{\rm opt}$ of a few photons. For $n =1$, the same time and $N_\pi$ are sufficient for $1-D_n < 0.01$ at a very small $\kappa/\chi \sim 0.1$. 

For a realistic threshold detector, it might be challenging to precisely choose $n_{\rm th} = n_{\rm opt}$ in Eq.~(\ref{detector-probability}). Moreover, in general, the probability to click deviates from an exact step function~(\ref{detector-probability}), which can be modeled by having a fluctuating
$n_{\rm th}$. This also applies to a perfect detector, but with an imperfect photon transfer. When the separation of the two distributions is large as in the continuous protocol of the previous section, $1-D_n$ is small for multiple values of $n$ around a large $n_{\rm opt}$. In this case, $n_{\rm th}$ does not need to be exactly $n_{\rm opt}$ and small fluctuations of $n_{\rm th}$ do not substantially increase the measurement error. In the sequential protocol, when $n_{\rm opt}$ is at most a few photons, even a one-photon deviation, i.e.,  $|n_{\rm th} -n_{\rm opt} | = 1$, may increase the measurement error above a tolerable threshold. To address this problem, one can  choose a range of parameters where $1-D_n$ is sufficiently small for several values of $n$ around $n=n_{\rm opt}$. For $N_\pi = 10$, we see that  the choice of $t = 3t_\pi$ and  $\kappa \sim 0.5\chi$ allows fluctuations of $n_{\rm th}$ between 2 and 3 photons while keeping $1-D_{n_{\rm th}}<0.01$.  If a wider range of allowable values of  $n_{\rm th}$ is required, it is necessary to choose a larger $N_\pi$.

\section{Beyond the dispersive limit}\label{sec-realistic}

In this section, we go beyond the dispersive limit and use the full-counting-statistics formalism to compute photon statistics for realistic parameters of a transmon qubit coupled to a cavity. In comparison to the previous section, when only two system parameters ($\kappa$ and $\chi$) were important, the relevant-parameter space is much larger beyond the dispersive limit. As such, the primary goal of this section is to illustrate important physical processes affecting distributions of emitted photons without scanning the entire parameter space.
We present calculations for a specific choice of frequencies and coupling constants characterizing the qubit-cavity system. For this choice, we study how photon statistics are affected by the cavity decay rate, the drive power, and the counting time.

We present our model in Sec.~\ref{sec-realistic-model} with details of the full-counting-statistics formalism given in Appendix~\ref{sec-fcs}. We discuss parameters for numerical simulations in Sec.~\ref{sec-realistic-simulations}. In Sec.~\ref{sec-realistic-main-effects}, we illustrate how the cavity state is affected by the cavity nonlinearity and the Purcell decay, which are two main effects affecting such state that were absent in the dispersive model of the previous section. In Secs.~\ref{sec-realistic-alwon} and \ref{sec-realistic-onoff}, we present and discuss emitted-photon statistics for the two protocols of photon counting.

\subsection{Model}\label{sec-realistic-model}

To focus on the main ideas, we assume the following simplifications in the model. We consider a transmon qubit~\cite{Koch2007} approximated by a weakly nonlinear Duffing oscillator with a constant negative anharmonicity $(-\eta)$ and invoke the rotating-wave approximation (RWA). Additionally, we work in the frame rotating at the drive frequency $\omega_d$. The unitary dynamics of such a cavity-qubit system is described by the Hamiltonian
\begin{equation}\label{Hamiltonian-general}
 \hat{H} = \hat{H}_{\rm cav} + \hat{H}_q + \hat{H}_{\rm int} + \hat{H}_{\rm drive}\,.
\end{equation}
In this expression, $\hat{H}_{\rm cav}$ is the isolated-cavity Hamiltonian, which is given by
\begin{equation}
 \hat{H}_{\rm cav} = \hbar(\omega_c - \omega_d)\left( \ha^\dagger \ha + 1/2\right )\,,
\end{equation}
where $\omega_c$ is the bare frequency of the relevant cavity mode. 
The multi-level transmon circuit is described by
\begin{equation}
 \hat{H}_q = \sum_{k\ge 0} (\varepsilon_k - k\hbar \omega_d) |k\rangle \langle k|\,,
\end{equation}
where $k$ labels its eigenstates and 
\begin{equation}
 \varepsilon_k = k\hbar\omega_q - k(k-1)\hbar\eta / 2
\end{equation}
is the energy of state $|k\rangle$. Here the bare 0-1 qubit transition frequency is $\hbar\omega_q \approx \sqrt{8E_C E_J}$ and the anharmonicity is $-\hbar\eta \approx - E_C$, where $E_J$ and $E_C$ are the Josephson and charging energies of the transmon ($E_J \gg E_C$)~\cite{Koch2007}. In the RWA, the qubit-cavity interaction is given by the usual Jaynes-Cummings form
\begin{equation}
 \hat{H}_{\rm int} = \hbar g \left(\ha^\dagger \hat{b} + \ha \hat{b}^\dagger\right)\,,
\end{equation}
where $g$ is the interaction strength and 
\begin{equation}
 \hat{b} = \sum_{k >0} \sqrt{k} |k-1\rangle\langle k|
\end{equation}
is the annihilation operator for the transmon. The last term of Eq.~(\ref{Hamiltonian-general}) describes coupling to a classical microwave drive:
\begin{equation}\label{Hamiltonian-drive}
 \hat{H}_{\rm drive} = \hbar \varepsilon(t) \left(\ha  + \ha^\dagger\right)
\end{equation}

For nonunitary dynamics, the cavity decay channel with the rate $\kappa$ is described by the Lindblad superoperator $\hat{L}_\kappa$ defined via its effect on the density matrix $\rho$:
\begin{equation}\label{L_kappa}
 \hat{L}_\kappa \rho = \frac{\kappa}{2} \left(2\hat{a} \rho\hat{a}^\dagger - \rho \hat{a}^\dagger\hat{a} -\hat{a}^\dagger\hat{a}\rho \right)\,. 
\end{equation}
We ignore any intrinsic qubit relaxation and dephasing, which is reasonable since typical intrinsic qubit lifetimes nowadays are much longer than time scales considered in this paper. Thus, we consider qubit relaxation that is coming only from the Purcell effect~\cite{Blais2004,  Sete2014}. 

 Let $\Delta = \omega_q - \omega_c$ be the detuning between the bare qubit and cavity frequencies. For the model of a weakly anharmonic transmon qubit described above, the $\chi$-shift (\ref{chi-definition}) in second-order perturbation theory, which is valid at $g\ll {\rm min}(|\Delta|, |\Delta - \eta|)$, is given by~\cite{Koch2007} ($\eta > 0$)
\begin{equation}\label{chi_expression}
 \chi = \frac{g^2}{\Delta} \frac{\eta}{\Delta - \eta}\,.
\end{equation}
We assume that our system is outside of the straddling regime~\cite{Koch2007, Boissonneault2012}, i.e., $\omega_c$ is outside of the interval ${\omega_q - \eta < \omega_c < \omega_q}$. This implies that $\omega_c^{(0)} > \omega_c^{(1)}$ and $\chi >0$ in agreement with Figs.~\ref{fig-htrdn-vs-jpm}(a) and \ref{fig-htrdn-vs-jpm}(d). With increasing the cavity occupation, the resonator becomes nonlinear and these dressed frequencies change and eventually approach the bare cavity frequency $\omega_c$~\cite{Boissonneault2010, Reed2010_prl, Khezri2016}. 

We calculate the distributions of emitted photons $P_n^{b}$ and $P_n^{d}$ using the full-counting-statistics formalism based on the quantum-jump approach~\cite{Zoller1987, Carmichael2008_book, Gardiner2008_book, Xu2013, Brange2019}, which is described in Appendix~\ref{sec-fcs}. 

\subsection{Numerical simulations}\label{sec-realistic-simulations}

We perform simulations for the following realistic parameters of the transmon qubit and cavity. The bare cavity and qubit frequencies are $\omega_c/ 2\pi = 5$ GHz and $\omega_q/2\pi = 4.5$ GHz, the qubit anharmonicity is $-\eta/2\pi = -250$ MHz, and the qubit-cavity coupling constant is $g/2\pi = 100$ MHz. This choice of parameters results in $\omega_c < \omega_c^{(1)} < \omega_c^{(0)}$ and in  $t_\pi = \pi/\chi = 75$ ns, where the $\chi$-shift is given by the perturbative expression (\ref{chi_expression}).
The motivation behind this choice is the proximity of parameters to experimental values of Ref.~\cite{Opremcak2018} and a relatively low effective critical number $n_{\rm crit} = \Delta^2/(4g^2) = 6.25$, which requires a smaller cavity occupation and, therefore, a smaller Hilbert space to probe the cavity nonlinearity in the simulations.

Our numerical procedure is as follows. In each simulation, we solved the master equation for five levels in the transmon and for the number of cavity levels chosen to be 2--2.5 times larger than the maximal cavity occupation in that particular simulation. 
In the full-counting-statistics calculations, see Appendix~\ref{sec-fcs}, we solved the master equation for an effective Lindblad operator at different values of the so-called counting field $\xi$ in the interval between 0 and $\pi$ and integrated numerically over $\xi$. 
Depending on $\overline{n}_b$, here the number of different values of $\xi$ varied between 250 and 1500.

\subsection{Cavity nonlinearity and Purcell decay}\label{sec-realistic-main-effects}

\begin{figure}[t]
 \includegraphics[width=0.95\columnwidth]{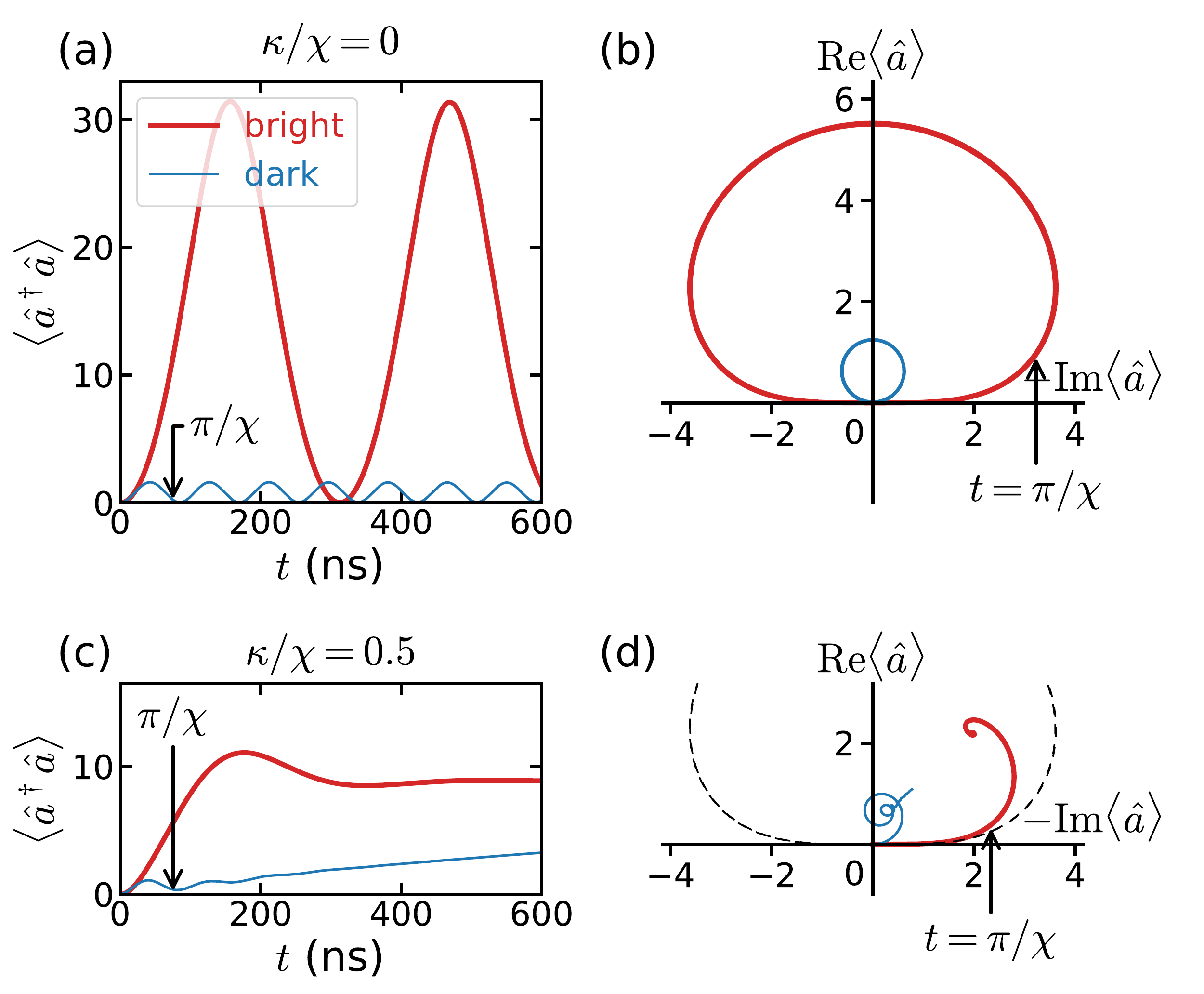}
 \caption{Cavity evolution beyond the dispersive limit at $\kappa/\chi = 0$ (a), (b) and at $\kappa/\chi = 0.5$ (c), (d). System parameters (see text) are chosen in such a way that $\omega_c < \omega_c^{(1)} < \omega_c^{(0)}$, $t_\pi = \pi/\chi = 75$ ns, and $n_{\rm crit} = \Delta^2/(4g^2) = 6.25$. The microwave drive is applied at $\omega_d=\omega_c^{(0)}$ with $(4\varepsilon / \chi)^2 = 20$. The dashed line of panel (d) shows the bright-state curve of panel (b) for comparison.
 }\label{fig-cav-tdep}
 \end{figure}

 \begin{figure}[t]
 \includegraphics[width=0.95\columnwidth]{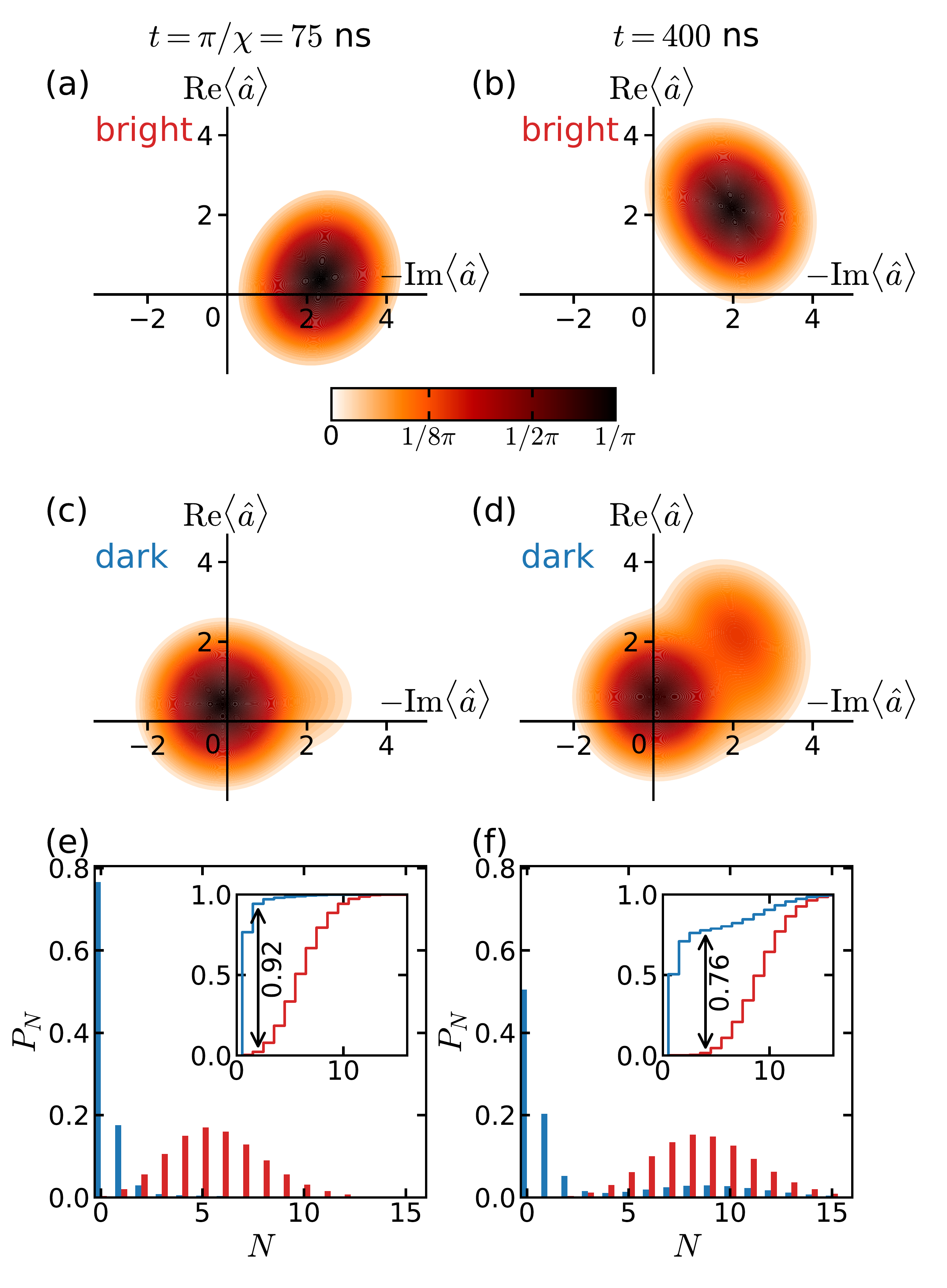}
 \caption{(a)--(d) Husimi $Q$ functions of the cavity calculated for the same parameters as in Fig.~\ref{fig-cav-tdep}(d) for the bright (a), (b) and dark (c), (d) states at $t=t_\pi = 75$ ns (a), (c) and $t=400$ ns (b), (d). (e), (f) The distributions of cavity photons for the same values of $t$ as the $Q$ functions. Insets: integrated distributions with the optimal separation thresholds and corresponding distances.}\label{fig-qfunc}
 \end{figure}
 
In the model of Sec.~\ref{sec-dispersive}, the cavity state has always been a coherent state. Here we elaborate on the major differences between such perfect coherent states and realistic bright and dark cavity states. These differences can be explained by the Purcell effect and cavity nonlinearity.

We start from briefly discussing these effects in the average metrics of the cavity state, which are illustrated in Fig.~\ref{fig-cav-tdep} for $\kappa/\chi =0$ [(a), (b)] and $\kappa/\chi = 0.5$ [(c), (d)]. 
As there is no Purcell decay at $\kappa = 0$, Figs.~\ref{fig-cav-tdep}(a) and \ref{fig-cav-tdep}(b) illustrate the effects of the cavity nonlinearity. There, instead of being a straight line as in Fig.~\ref{fig-htrdn-vs-jpm}(e), the bright-state trajectory in the phase space is a closed curve, so $\overline{N}_b(t)$ is periodic with maxima at the topmost points, which satisfy ${\rm Im}[\alpha_b(t)] = 0$. Since there is no Purcell decay, this behavior can still be understood qualitatively by extending the semiclassical formalism for coherent states of the previous section. Assuming the cavity resonance frequency $\omega_c^{(0/1)}$ now depends on the photon number $N$ and is given by the difference between the corresponding eigenenergies of the Hamiltonian (\ref{Hamiltonian-general}) with $N$ and $N+1$ photons,  we introduce this $N$ dependence into the cavity-drive detuning (\ref{detuning-cavity-drive}) and analytically continue this detuning $\Delta_{{\rm cd}}(N)$ defined at integer $N$ into real values of $|\alpha|^2$. For the lowest-order or Kerr nonlinearity, 
\begin{equation}\label{detuning-kerr}
\Delta_{{\rm cd}}\left(|\alpha|^2\right) = -\zeta |\alpha|^2
\end{equation}
in the bright state, where, for our choice of parameters, $\zeta > 0$.
We then rewrite Eq.~(\ref{alpha-semiclassical-eom}) at $\kappa = 0$ as
\begin{equation}
 \dot{\alpha} = -i\Delta_{{\rm cd}}\left(|\alpha|^2\right)\alpha - i\varepsilon\,.
\end{equation}
Since the cavity state is no longer coherent, $|\alpha|^2$ is no longer strictly $\overline{N}$ and this equation is no longer exact, so we use it only for qualitative understanding. 

Introducing angle $\varphi$ as $\alpha = -i |\alpha| e^{i\varphi}$, which is the polar angle in the convention of phase-space figures of this paper, which follows $(-{\rm Im}\alpha, {\rm Re} \alpha)$ coordinates, we find
\begin{eqnarray}
 \frac{d|\alpha|}{dt} &=& \varepsilon \cos\varphi\,,\\
 |\alpha| \frac{d\varphi}{dt} &=& - \Delta_{{\rm cd}}\left(|\alpha|^2\right)|\alpha| - \varepsilon \sin \varphi\,.
\end{eqnarray}
Thus, it is clear from the first equation that the maxima of the cavity occupation at Fig.~\ref{fig-cav-tdep}(a) correspond to $\varphi = \pi/2$ in Fig.~\ref{fig-cav-tdep}(b). For the Kerr-type nonlinearity (\ref{detuning-kerr}), it is possible to write an analytic expression for  the bright-state phase-space curve:
\begin{equation}\label{trajectory-kerr}
 |\alpha|^3 = \frac{4\varepsilon}{\zeta}\sin\varphi\,.
\end{equation}
On a qualitative level, this expression explains the shape of the trajectory in Fig.~\ref{fig-cav-tdep}(b).

The bright-state trajectories  of Figs.~\ref{fig-cav-tdep}(c) and (d), calculated at $\kappa \ne 0$, further highlight nonlinear effects. In panel (c), the steady-state occupation is slightly less than 10 photons even though $(2\varepsilon/\kappa)^2 = 20$, which would have  resulted in $\overline{N}_b(t\rightarrow \infty) = 20$ in the dispersive approximation; see Eq.~(\ref{ncav-t-disp}). Similar to the $\kappa=0$ case, [Fig.~\ref{fig-cav-tdep}(b) and  Eq.~(\ref{trajectory-kerr})], the phase-space trajectory is no longer a straight line. Unlike the dispersive case, the steady state is located off the $x$ axis. In addition to the cavity nonlinearity, these figures demonstrate another important effect of realistic systems: the Purcell decay. Since we have chosen $\omega_d = \omega_c^{(0)}$, this effect is visible in the dark-state trajectories, which correspond to qubit in its excited state. One can see that the average cavity occupation in the dark state increases with time as qubit relaxes to its ground state, which effectively changes the cavity resonance frequency. In the phase space, there is only one global steady state solution, which is the bright steady state for the chosen drive frequency. Thus, the dark-state trajectory in Fig.~\ref{fig-cav-tdep}(d) slowly converges towards the bright steady state.

One consequence of both cavity nonlinearity and Purcell decay is that the cavity state is no longer coherent. We illustrate this by plotting the Husimi $Q$ function in Fig.~\ref{fig-qfunc} corresponding to two points of each trajectory of Fig.~\ref{fig-cav-tdep}(d). Namely, we do it for $t=\pi/\chi = 75$ ns, when the dark-state $\overline{N}_d$ is close to its minimum, and for $t=400$ ns, when the bright steady state has been reached. The $Q$ function is a function of the phase-space complex coordinate $\beta$ and is defined as 
\begin{equation}
Q(\beta) = \frac{1}{\pi}\langle \beta |\rho_{\rm cav}|\beta\rangle\,.
\end{equation}
Here $|\beta\rangle$ is the corresponding coherent state, and $\rho_{\rm cav}$ is the reduced cavity density matrix for a given system density matrix $\rho$. In addition to this function, we show the corresponding cavity photon distributions at the bottom of Fig.~\ref{fig-qfunc}. These distributions are obtained from diagonal elements of $\rho_{\rm cav}$. For each pair of bright and dark cavity photon distributions, we can define the distance between them the same way we defined it for the distributions of emitted photons in Eq.~(\ref{KS-definition}). We plot the corresponding cumulative distributions in the insets of Figs.~\ref{fig-qfunc}(e) and \ref{fig-qfunc}(f) and show such distances.

Since there is no Purcell decay for the bright state, we observe that the maxima of two $Q$ functions in Figs.~\ref{fig-qfunc}(a) and \ref{fig-qfunc}(b) are very close to the corresponding points of $\alpha_b(t)$ in Fig.~\ref{fig-cav-tdep}(d). However, the shape of those $Q$ functions implies that the states deviate from being coherent. At $t=400$ ns, the state is squeezed and the uncertainty in photon number is slightly reduced. This is explained by the dependence of the cavity resonance frequency on the photon number, so, for the bright state, the points in the phase space that are further away from the origin have higher angular speed~\cite{Khezri2016}. We notice that the squeezing of the cavity state can potentially benefit photon-detector-based readout since it reduces the photon-number uncertainty.

The $Q$ function for the dark state is more involved. First, its deformed shape demonstrates the Purcell effect. At $t=400$ ns, it is more asymmetric and it shows how the dark quasi steady state slowly transitions towards the global bright steady state. Second, one can see that the maximum of the $Q$ function no longer coincides with the corresponding average value $\langle \hat{a}\rangle$; i.e., the maximum in Fig.~\ref{fig-qfunc}(d) is much closer to the $y$ axis than the corresponding point in the blue thin line of Fig.~\ref{fig-cav-tdep}(d). Figuress~\ref{fig-qfunc}(e) and \ref{fig-qfunc}(f) illustrate that, because of this Purcell effect, the dark-state cavity-photon distribution acquires a tail at large photon numbers, which increases at longer times. This is especially evident from the inset, where the optimal distance between two cumulative distributions decreases with time. A possible way to circumvent this problem in a readout experiment is to use a Purcell filter~\cite{Reed2010_apl, Jeffrey2014, Sete2015, Bronn2015}. However, due to the intrinsic qubit relaxation, a similar deformation of the photon distribution function will remain.
 
 \subsection{Continuous counting}\label{sec-realistic-alwon}

\begin{figure}[t]
 \includegraphics[width=0.95\columnwidth]{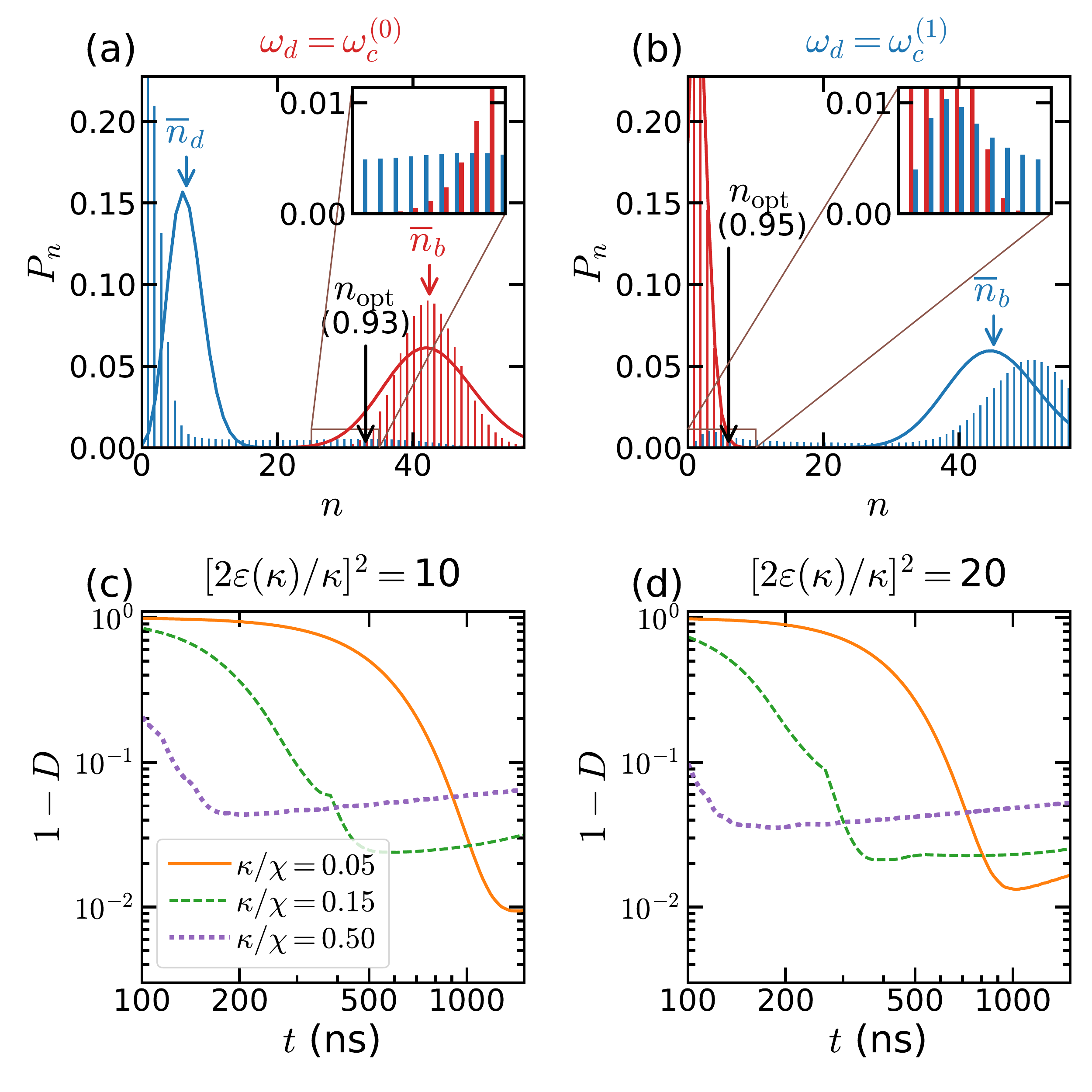}
 \caption{Continuous photon counting in a realistic system. (a), (b) Emitted-photon distributions $P_n^b$ and $P_n^d$ (vertical bars) for $\kappa/\chi = 0.5$, $t=400$ ns, $(2\varepsilon/\kappa)^2 = 10$ with the microwave drive applied at $\omega_c^{(0)}$ (a) and $\omega_c^{(1)}$ (b). Solid lines show the Poisson distributions calculated for the  mean values $\overline{n}_b$ and $\overline{n}_d$ of realistic $P_n^b$ and $P_n^d$. The optimal thresholds $n_{\rm opt}$ are shown by vertical arrows with distances $D$ written in brackets. (c), (d) Distribution overlaps $1-D$ vs $t$ calculated for $\omega_d = \omega_c^{(1)}$ at three different values of $\kappa/\chi$ shown in the legend. The drive power is scaled with $\kappa$  to satisfy $(2\varepsilon/\kappa)^2 = 10$ (c) and $20$ (d).}\label{fig-fid-alw-on}
\end{figure}

We now discuss the statistics of emitted photons in the continuous counting protocol, when the application of microwave drive and photon counting happen at the same time.

First, in Fig.~\ref{fig-fid-alw-on}(a), we show $P_n^b$ and $P_n^d$ for the same value of the decay rate $\kappa/\chi = 0.5$ and the same drive frequency $\omega_d = \omega_c^{(0)}$ as in Figs.~\ref{fig-cav-tdep}(c) and \ref{fig-cav-tdep}(d) and Fig.~\ref{fig-qfunc}. To visually improve the figure, we do it for the smaller value of the drive power satisfying $(2\varepsilon / \kappa)^2 = 10$. The distributions are calculated at $t=400$ ns, which corresponds to the right column of Fig.~\ref{fig-qfunc}. A tail that is caused by qubit relaxation  is present in $P_n^d$ at large $n$, which is clearly visible in the inset showing details at larger scale. To further illustrate the effects of the nonlinearity and the Purcell decay, we look at $\overline{n}_b$ and $\overline{n}_d$, the mean values of realistic $P_n^b$ and $P_n^d$, and plot the corresponding Poissonian distributions $P_{\rm Poiss}(\overline{n}_b)$ and $P_{\rm Poiss}(\overline{n}_d)$ for these mean values (solid lines). Because $P_n^d$ has a tail, $\overline{n}_d > n_{\rm max}^d$, where $n_{\rm max}^i$ is the value of $n$ at which $P_n^i$ has its maximum ($i=b, d$). The second observation is that, while $\overline{n}_b\approx n_{\rm max}^b$, the bright distribution is more narrow than $P_{\rm Poiss}(\overline{n}_b)$, which is explained by  squeezing of the cavity state discussed above. 

In Fig.~\ref{fig-fid-alw-on}(b), we show these distributions for the same parameters as Fig.~\ref{fig-fid-alw-on}(a), but for the microwave drive applied at $\omega_c^{(1)}$ instead of $\omega_c^{(0)}$. In this case, it is $P_n^b$ that displays the effect of qubit relaxation with the tail being at small  $n$. As a consequence, $P_{\rm Poiss}(\overline{n}_b)$ is now shifted to smaller values of $n$. In comparison to Fig.~\ref{fig-fid-alw-on}(a), $n_{\rm  max}^b$ is now larger ($\sim 55$ vs $\sim 40$ photons), and squeezing effects are much less noticeable. Both of these facts are explained by a smaller degree of resonator nonlinearity for qubit in state 1, which is consistent with $|\omega_c^{(1)} - \omega_c| < |\omega_c^{(0)} - \omega_c|$ for our choice of parameters. 

Because of the long tails, the optimal threshold $n_{\rm opt}$ is far from the middle between $n_{\rm max}^b$ and $n_{\rm max}^d$ in both cases. Because of different long tails, $n_{\rm opt}$ in Fig.~\ref{fig-fid-alw-on}(b) is much smaller than $n_{\rm opt}$ in Fig.~\ref{fig-fid-alw-on}(a). In a readout experiment with a photon detector, this may make  $\omega_d = \omega_c^{(1)}$ to be a better choice than $\omega_d = \omega_c^{(0)}$ if a smaller threshold is preferable. In addition, for a fixed drive amplitude, weaker nonlinearity makes the distance $D$ somewhat larger at $\omega_d = \omega_c^{(1)}$ (0.95 vs 0.93 for the example under consideration).

In Figs.~\ref{fig-fid-alw-on}(c) and \ref{fig-fid-alw-on}(d), we plot the  overlap $1-D$ vs counting time $t$ for $\omega_d = \omega_c^{(1)}$ and three values of $\kappa/\chi$. For each value of $\kappa/\chi$, we choose the drive amplitude $\varepsilon(\kappa)$ to ensure  $(2\varepsilon / \kappa)^2 = 10$ in Fig.~\ref{fig-fid-alw-on}(c) and $(2\varepsilon / \kappa)^2 = 20$ in Fig.~\ref{fig-fid-alw-on}(d). Thus, in each of these figures, the maximal bright-state cavity occupation is approximately the same for different values of $\kappa$. The $\kappa/\chi = 0.5$ curve of Fig.~\ref{fig-fid-alw-on}(c) corresponds to Fig.~\ref{fig-fid-alw-on}(b). 

The first observation from Figs.~\ref{fig-fid-alw-on}(c) and \ref{fig-fid-alw-on}(d) is that, for each $\varepsilon$ and $\kappa$, there is a local minimum in $1-D$ vs $t$. This effect is qualitatively different from the monotonic behavior in the dispersive model; see Fig.~\ref{fig-fid-disp-alw-on}(c). It is explained by the competition between two processes: increasing numbers of emitted photons with $t$ and qubit relaxation. At shorter times, the overlap is determined by an increasing distance between $\overline{n}_b$ and $\overline{n}_d$, which is similar to the dispersive model.
At longer time, the overlap is mostly determined by  the left tail of $P_n^b$ rather than by the distance between $\overline{n}_b$ and $\overline{n}_d$, which reduces the overlap with time. 

The second observation is that the optimal $t$ and the corresponding maximal value of $D$ increase with decreasing $\kappa/\chi$. The shift of the optimal $t$ to longer times is explained by slowing down of all the relevant rates (i.e., both Purcell decay and photon emission). The reduction in $1-D$ can be explained as follows. Let us fix some value of $\overline{n}_b$. The time required to achieve this value of $\overline{n}_b$, apparently, increases with decreasing $\kappa/\chi$. Since the Purcell decay and the rate of growth of  $\overline{n}_b$ slow down by approximately the same factor with reducing $\kappa/\chi$, the left tail of the bright distribution would not change appreciably with reducing $\kappa/\chi$ for a fixed $\overline{n}_b$. However, $\overline{N}_d$ is smaller at smaller $\kappa/\chi$ for the same $\overline{N}_b$, see Eq.~(\ref{ncav-t-disp}). Therefore, the overlap between the dark-state distribution and the tail of the bright-state distribution decreases.

Another interesting effect is that increasing the drive power or the cavity occupation is not very effective in reducing $1-D$; compare Figs.~\ref{fig-fid-alw-on}(c) and \ref{fig-fid-alw-on}(d). This is well explained by the cavity nonlinearity, which increases with increasing the cavity occupation. Thus, while the dark-state $\overline{N}_d$ approximately scales with the drive power as $\varepsilon^2$ and, therefore, increases twofold between the two figures for the same value of $\kappa/\chi$, the bright-state $\overline{N}_b$ is larger and, therefore, grows slower than $\varepsilon^2$. Overall, this slows the expected growth of separation between the bright and dark distributions. In some cases such as $\kappa/\chi=0.5$, increasing $\varepsilon$ can actually increase the overlap.

\subsection{Sequential counting}\label{sec-realistic-onoff}

\begin{figure}[t]
 \includegraphics[width=0.95\columnwidth]{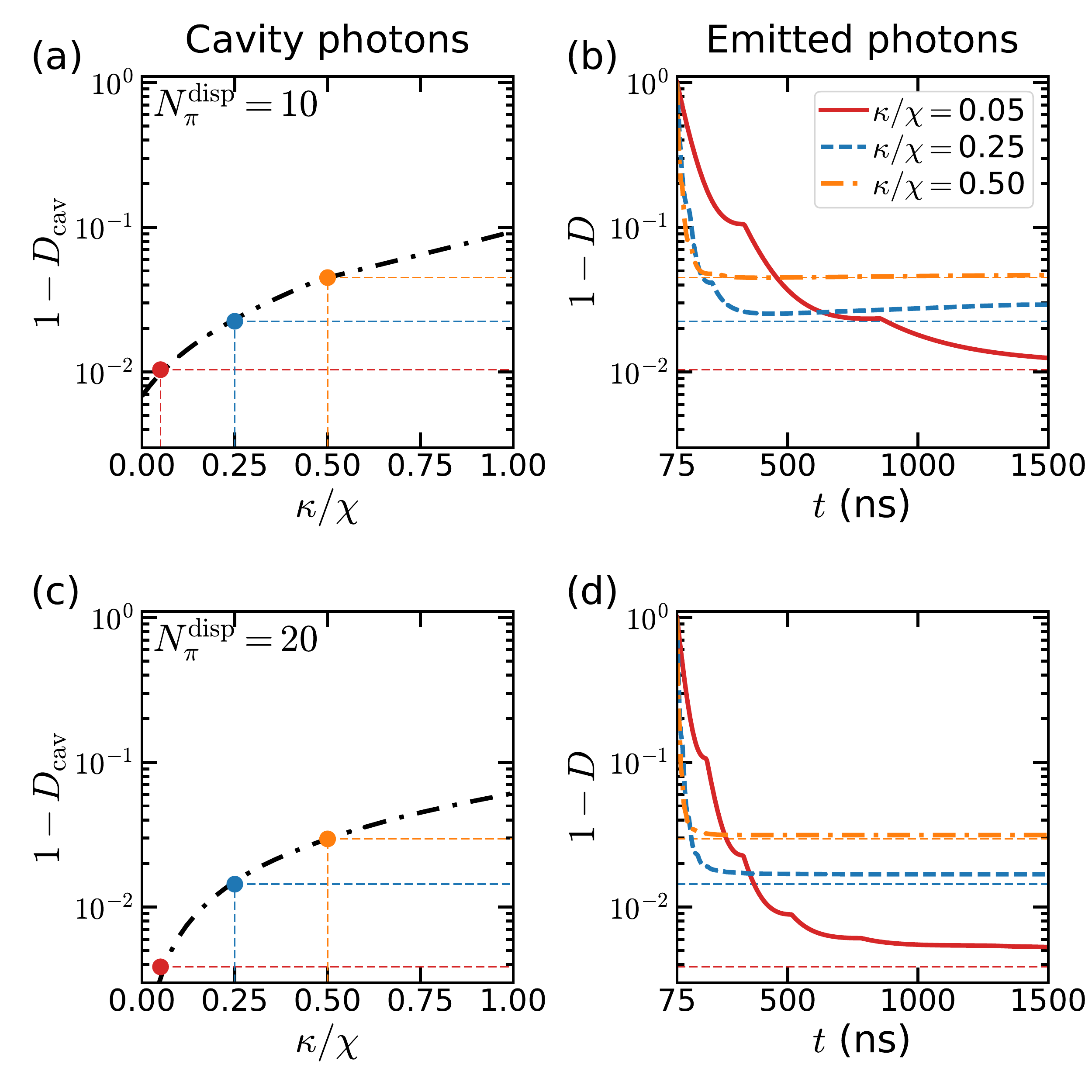}
 \caption{Sequential counting  in a realistic system. (a), (c) The overlap $1-D_{\rm cav}$ between photon distributions of cavity pointer states at $t=t_\pi = 75$ ns as a function of $\kappa/\chi$ at $\varepsilon(\kappa)$ satisfying $N_\pi^{\rm disp}(\varepsilon, \kappa) = 10$ (a) and $N_\pi^{\rm disp}(\varepsilon, \kappa) = 20$ (c), where $N_\pi^{\rm disp}(\varepsilon, \kappa)$ is given by $N_\pi$ of Eq.~(\ref{ncav-bright-t-pichi}). 
 (b), (d) $1-D$ vs $t$ at $t\ge t_\pi$ for three values of $\kappa/\chi$ and the corresponding values of $N_\pi^{\rm disp}$ from the left column. Three values of $\kappa/\chi$ are shown by vertical lines in  the left column, and the corresponding values of $1-D_{\rm cav}$ are shown by horizontal lines in all the panels. }\label{fig-fid-onoff}
\end{figure}

We now discuss the second protocol of photon counting, which was defined and discussed in Sec.~\ref{sec-dispersive-onoff} and Fig.~\ref{fig-fid-disp-onoff} for the dispersive approximation. The main results of this protocol for a realistic system are presented in Fig.~\ref{fig-fid-onoff} for $\omega_d = \omega_c^{(0)}$.

Since the cavity states are prepared before photon counting, the distance $D$ is limited by the quality of the pointer states. As a measure characterizing this quality, we consider the distance between photon distributions of cavity pointer states $D_{\rm cav}$, which we define exactly the same way as the distance (\ref{KS-definition}) but for the cavity photon distributions $P_N$. Such distributions and the corresponding distance were shown in Figs.~\ref{fig-qfunc}(e) and \ref{fig-qfunc}(f). We notice that $D_{\rm cav}$ characterizes the resolvability of photon distributions; it is not a measure  of how well the pointer states are separated in general since information about the phase is lost in photon distributions.

In the left column of Fig.~\ref{fig-fid-onoff}, we show  $1 - D_{\rm cav}$ as a function of $\kappa/\chi$. Similar to Fig.~\ref{fig-fid-disp-onoff}, we show results vs $\kappa$ at a fixed bright-state pointer occupation $N_\pi$ rather than $\varepsilon$. For a realistic system with cavity nonlinearity, we do it only approximately and fix the expected value of $N_\pi$ using the dispersive relation (\ref{ncav-bright-t-pichi}). Thus, for each value of $\kappa/\chi$, we calculate $\varepsilon$ using (\ref{ncav-bright-t-pichi}) with this value of dispersive $N_\pi$ being 10 [panel (a)] and 20 [panel (c)]. Both panels of the left column demonstrate that $1-D_{\rm cav}$ decreases with $\kappa/\chi$ decreasing because the latter improves the dark pointer state by reducing $\overline{N}_d(t_\pi)$ and suppressing the Purcell tail. The value of $1-D_{\rm cav}$ at $\kappa/\chi = 0$ gives the theoretical limit on the quality of pointer states for a given number of photons in the bright state. In relation to qubit measurements with an ideal photon counter, we therefore see that, for $N_\pi \approx 10$, the best possible measurement fidelity can only slightly exceed 99\%, while at $N_\pi \approx 20$ it can go above 99.9\%. 

In the right column of Fig.~\ref{fig-fid-onoff}, we show $1 - D$ vs time for three values of $\kappa/\chi$ for the same values of the $\kappa$-dependent drive power as in the corresponding panels of the left column. For this reason, we highlight with circles the points in $1-D_{\rm cav}$ vs $\kappa$ curves of the left column for these three values of $\kappa/\chi$ and plot horizontal lines accordingly as a guide to the eye. Thus, in the right column, we observe that $1 - D$ approaches but does not touch these horizontal lines, which illustrates the impossibility to resolve $P_n^b$ and $P_n^d$ better than the resolution between cavity pointer states. However, $1 - D$ does not strictly approach $1-D_{\rm cav}$ asymptotically at $t\rightarrow \infty$, but has a minimum, which can be seen in the dashed line of Fig.~\ref{fig-fid-onoff}(b). This is explained by qubit transitions at $t> t_\pi$ and is characteristic for our choice of the drive frequency. Since $\omega_d = \omega_c^{(0)}$, the tail of $P^{d}_n$ can increase in comparison to the tail in the pointer-state distribution because qubit in its excited state can relax and, thus, emit an extra photon into the transmission line even at $t>t_\pi$.

\section{Conclusions}\label{sec-conclusions}

We have studied theoretically the statistics of photons emitted by a microwave resonator in the circuit QED architecture. Motivated by qubit measurements, we focused on differentiating between such statistics for the two qubit states. We performed calculations for the case of bright and dark cavity pointer states, which ensure  distinct photon occupations for the two qubit states. Our quantitative results are applicable to qubit measurements using an ideal photon counter or threshold photodetector.

We studied two protocols of photon counting based on whether the application of a microwave drive and photon counting occur simultaneously or one after another.
We found that, in the dispersive approximation and at negligible qubit relaxation, the simplest continuous protocol (simultaneous driving and counting) allows 1\% overlap between two distributions for cavity occupations not exceeding 10 photons within a very short time. This time is of the order of $t_\pi$, which is the minimal time required by the time-frequency uncertainty relation to resolve two dressed cavity frequencies and, therefore, to perform qubit measurement.
A time that is twice as long results in the minuscule overlap error of less than 0.01\% without any fine-tuning of the cavity decay rate. The sequential 
protocol, in which pointer-state preparation and photon counting are separated in time, gives larger overlap errors. For optimal values of the decay rate, the error in the sequential protocol can still be reduced below 1\% for the same maximum cavity occupation of about 10 photons and the combined drive and counting time of $2t_\pi$.
This optimal decay rate, which minimizes the distribution overlap, is parametrically smaller than the one optimizing the signal-to-noise ratio in heterodyne technique.

For a realistic system, we demonstrated that the resolvability between two distributions is reduced significantly because of the qubit relaxation. 
For both protocols of photon counting, the distributions overlap below 1\% can be achieved by using a system with a very small decay rate and a relatively long counting time, which is an order of magnitude longer than $t_\pi$. For the same maximum cavity occupation, this overlap is smaller in the sequential protocol, when the error due to Purcell decay is smaller. 

Our quantitative results for a realistic system are applicable for a specific parameter choice of the transmon qubit~\cite{Koch2007}. Other qubit designs  with a longer lifetime~\cite{Nguyen2019} can improve the separation between distributions. However, this separation is likely bounded by the results of the dispersive approximation. Another way to reduce both the overlap and the  counting  time in the sequential protocol is to use a catch-and-release-like method~\cite{Yin2013, Sete2013, Peronnin2019} provided a resonator with a tunable decay rate is available. 
A Purcell filter~\cite{Reed2010_apl, Jeffrey2014, Sete2015, Bronn2015} can also provide a way to reduce the overlap error.

\acknowledgments
We are grateful to R. McDermott, A. Opremcak, B. Christensen, K. Kechedgi, and Z. Qi for fruitful discussions.  This work was supported by the U.S. Army Research Office (Grants No. W911NF-15-1-0248 and No. W911NF-18-1-0146) and NSF PFC at JQI (Grant No. 1430094). We acknowledge the use of the QuTiP software package~\cite{Johansson2012, Johansson2013}. Numerical simulations were performed using  the compute resources and assistance of the UW-Madison Center For High Throughput Computing (CHTC) in the Department of Computer Sciences. The CHTC is supported by UW-Madison, the Advanced Computing Initiative, the Wisconsin Alumni Research Foundation, the Wisconsin Institutes for Discovery, and the National Science Foundation, and is an active member of the Open Science Grid, which is supported by the National Science Foundation and the U.S. Department of Energy's Office of Science.

\appendix

\section{Full counting statistics}\label{sec-fcs}

Here we briefly describe the full-counting-statistics technique based on the quantum jump approach~\cite{Zoller1987, Carmichael2008_book, Gardiner2008_book, Xu2013, Brange2019}. 

The density matrix $\rho(t)$ satisfies the master equation
\begin{equation}\label{master-equation}
 \dot{\rho}(t) = \hat{\cal L}\rho(t) = -\frac{i}{\hbar}\left[\hat{H}(t), \rho(t)\right] + \hat{L}_\kappa \rho(t)\,,
\end{equation}
where the Hamiltonian $\hat{H}(t)$ is given by Eq.~(\ref{Hamiltonian-general}) with a possible time dependence in the drive term (\ref{Hamiltonian-drive}), and the Lindblad superoperator describing the cavity leakage is given by Eq.~(\ref{L_kappa}). Let $t_0$ be the time at which the photon counting starts as in Eq.~(\ref{nbar-definition}). Then, at $t \ge t_0$, we can express the density matrix as
\begin{equation}\label{rho=sum-rho-n}
 \rho(t) = \sum_{n\ge 0} \rho^{(n)}(t)\,,
\end{equation}
where $\rho^{(n)}(t)$ describes the quantum trajectory corresponding to exactly $n$ photons being emitted by the cavity in the interval $(t_0, t)$. Therefore, the probability of such outcome is given by
\begin{equation}\label{P_n_fcs}
 P_n(t) = {\rm Tr}\, \rho^{(n)}(t)\,.
\end{equation}
The goal of the full counting statistics is to calculate this probability for an arbitrary $n$.

The equation of motion for $\rho^{(n)}$ is given by
\begin{equation}\label{master-equation-rho-n}
 \dot{\rho}^{(n)}(t) = \hat{\cal L}_0 \rho^{(n)}(t)  + \hat{\cal J} \rho^{(n-1)}(t)\,,
\end{equation}
where $\hat{\cal L}_0 = \hat{\cal L} - \hat{\cal J}$ and $\hat{\cal J}$ is the jump superoperator defined as $\hat{\cal J}\rho = \kappa \hat{a}\rho\hat{a}^\dagger$. This operator describes stochastic quantum jumps associated with photon emission and thus couples trajectories with $n-1$ and $n$ photons. 
The boundary conditions for the equations (\ref{master-equation-rho-n}) are $\rho^{(n)}(t_0) = 0$ for $n>0$ and $\rho^{(0)}(t_0) = \rho(t_0)$. 
To calculate the probability (\ref{P_n_fcs}), we first introduce the generalized density matrix
\begin{equation}\label{rho-generalized}
 \tilde{\rho}(t, \xi) = \sum_{n\ge 0} e^{i n\xi}\rho^{(n)}(t)\,,
\end{equation}
where $\xi \in [0, 2\pi)$. Using Eq.~(\ref{master-equation-rho-n}), we find that it satisfies the equation of motion
\begin{equation}\label{master-equation-rho-gen}
 \dot{\tilde{\rho}}(t, \xi) = (\hat{\cal L}_0 + e^{i\xi }\hat{\cal J}) \rho(t, \xi)
\end{equation}
with the boundary condition $\tilde{\rho}(t_0, \xi) = \rho(t_0)$. The generalized density matrix gives the generating function
\begin{equation}
 {\cal G}(t, \xi) = {\rm Tr}\, \tilde{\rho}(t, \xi)  = \sum_{n \ge 0} e^{in\xi} P_n(t)\,.
\end{equation}
The inverse Fourier transform of ${\cal G}(t, \xi)$ gives the probability to count $n$ photons:
\begin{equation}\label{P_n_integral}
 P_n(t) = \int \limits_0^{2\pi} {\cal G}(t, \xi) e^{-in\xi} \frac{d\xi}{2\pi}\,.
\end{equation}

In practice, we find ${\cal G}(t, \xi)$ by solving Eq.~(\ref{master-equation-rho-gen}) numerically for a number of discrete values of $\xi\in [0, \pi]$. In our simulations, this number varied between 250 and 1500 depending on the average number of emitted photons. Noticing that ${\cal G}(t, \xi) = {\cal G}^*(t, 2\pi - \xi)$, we find ${\cal G}(t, \xi)$ for $\xi \in (\pi, 2\pi)$ and calculate the integral (\ref{P_n_integral}) numerically.

\bibliography{literature_jpm}

\end{document}